\documentclass[aps,twocolumn,showpacs,preprintnumbers,floats]{revtex4}\def\@cite#1#2{\textsuperscript{[{#1\if@tempswa , #2\fi}]}}
\usepackage{mathrsfs}
\usepackage{longtable,lscape}
\usepackage{txfonts}
\usepackage{amssymb}
\usepackage{indentfirst}
\usepackage{graphicx,booktabs}
\usepackage{color}
\usepackage{amssymb}
\usepackage{epsfig}
\newcommand{\vsig}{\mbox{\boldmath$\sigma$\unboldmath}}

\begin{document}
\title{$\Xi$ baryon strong decays in a chiral quark model }
\author{
Li-Ye Xiao and Xian-Hui Zhong \footnote {zhongxh@hunnu.edu.cn}}
\affiliation{ Department of Physics, Hunan Normal University, and
Key Laboratory of Low-Dimensional Quantum Structures and Quantum
Control of Ministry of Education, Changsha 410081, China }


\begin{abstract}
The strong decays of $\Xi$ baryons up to $N=2$ shell are studied in
a chiral quark model. The strong decay properties of these
well-established ground decuplet baryons are reasonably described.
We find that (i) $\Xi(1690)$ and $\Xi(1820)$ could be assigned to
the spin-parity $J^P=1/2^-$ state
$|70,^{2}{8},1,1,\frac{1}{2}^-\rangle$ and the spin-parity
$J^P=3/2^-$ state $|70,^{2}{8},1,1,\frac{3}{2}^-\rangle$,
respectively. Slight configuration mixing might exist in these two
negative parity states. (ii) $\Xi(1950)$ might correspond to several
different $\Xi$ resonances. The broad states ($\Gamma\sim 100$ MeV)
observed in the $\Xi\pi$ channel could be classified as the pure
$J^P=5/2^-$ octet state $|70,^{4}8,1,1,\frac{5}{2}^-\rangle$ or the
mixed state $|\Xi \frac{1}{2}^-\rangle_3 $ with $J^P=1/2^-$. The
$\Xi$ resonances with moderate width ($\Gamma\sim 60$ MeV) observed
in the $\Xi\pi$ channel might correspond to the $J^P=1/2^+$
excitation $|56,^{4}10,2,2,\frac{1}{2}^+\rangle$. The second orbital
excitation $|56,^{4}10,2,2,\frac{3}{2}^+\rangle$ and the mixed state
$|\Xi \frac{1}{2}^-\rangle_1$ might be candidates for the narrow
width state observed in the $\Lambda \bar{K}$ channel, however,
their spin-parity numbers are incompatible with the moment analysis
of the data. (iii) $\Xi(2030)$ could not be assigned to either any
spin-parity $J^P=7/2^+$ states or any pure $J^P=5/2^+$
configurations. It seems to favor the
$|70,^{2}8,2,2,\frac{3}{2}^+\rangle$ assignment, however, its spin
conflicts with the moment analysis of the data. To find more $\Xi$
resonances, the observations in the $\Xi(1530)\pi$ and $\Sigma(1385)
\bar{K}$ channels are necessary.
\end{abstract}


\maketitle

\section{INTRODUCTION}

Understanding the baryon spectrum and the search for the missing
baryon resonances are hot topics in hadron physics. For the scarcity
of high quality antikaon beams and the small production rate for the
$\Xi$ resonances via an electromagnetic probe, the $\Xi$ spectrum is
still far from being established. There are only a few data on the
$\Xi$ resonances from the old bubble chamber experiments with small
statistics. Among the eleven $\Xi$ baryons listed in the review of
the Particle Data Group (PDG) ~\cite{Beringer:1900zz}, only the
ground octet and decuplet, $\Xi(1320)$ and $\Xi(1530)$, are well
established with four-star ratings. Although the resonances
$\Xi(1690)$, $\Xi(1820)$, $\Xi(1950)$ and $\Xi(2030)$ are three-star
states in PDG, only $\Xi(1820)$ has a determined spin-parity
$J^P=3/2^-$.

A few theoretical studies of the $\Xi$ baryon spectrum can be found
in the literature~\cite{Isgur:1978xj,Chao:1980em,
Capstick:1986bm,Glozman:1995fu,Bijker:2000gq,
Pervin:2007wa,Oh:2007cr,Melde:2008yr,Chen:2009de,Lee:2002jb,Schat:2001xr,Goity:2002pu,Goity:2003ab}.
All of the phenomenological models can well explain the two ground
states $\Xi(1320)$ and $\Xi(1530)$, however, their predictions for
the excitations are very different. For example, there is a lot of
controversy about the spin-parity of $\Xi(1690)$. It might be the
first radial excitation of $\Xi$ with $J^P=1/2^+$ according to the
mass calculations with a nonrelativistic quark model by Chao
\emph{et al.}~\cite{Chao:1980em}. This classification was also
supported by the recent quark model study of Melde \emph{et
al.}~\cite{Melde:2008yr}. However, with a relativized quark model
Capstick and Isgur predicted that the first radial excitation of
$\Xi$ should have a larger mass of $\sim 1840$
MeV~\cite{Capstick:1986bm}. Recently, Pervin and Roberts suggested
that $\Xi(1690)$ could be assigned to the first orbital excitation
of $\Xi$ with $J^P=1/2^-$, although their quark model predicted mass
is about 35 MeV heavier than the mass of
$\Xi(1690)$~\cite{Pervin:2007wa}. They believed that a more
microscopic treatment of spin-orbit interactions, can be expected to
drive this state to slightly lower mass. The calculations from the
Skyrme model also indicated that $\Xi(1690)$ should have
$J^P=1/2^-$~\cite{Oh:2007cr}. Furthermore, $\Xi(1690)$ was discussed
to be a dynamically generated state with
$J^P=1/2^-$~\cite{Kolomeitsev:2003kt,Sarkar:2004jh}. It should be
mentioned that \emph{BABAR} collaboration, in a study of
$\Lambda_c^+\rightarrow \Xi^-\pi^+K^+$, recently found some evidence
that $\Xi(1690)$ has $J^P=1/2^-$~\cite{Aubert:2008ty}.

Even for the $\Xi(1820)$ with well determined spin-parity quantum
numbers, $J^P=3/2^-$, there still exist many puzzles in the
explanations of its nature. Considering $\Xi(1820)$ as a $J^P=3/2^-$
orbital excitation, its mass and decay properties could be well
explained in the nonrelativistic quark model~\cite{Chao:1980em} and
algebraic model~\cite{Bijker:2000gq}. However, considering
$\Xi(1820)$ as a $J^P=3/2^-$ excitation, the mass of $\Xi(1820)$ is
obviously underestimated by the relativized quark
model~\cite{Capstick:1986bm} and one-boson exchange
model~\cite{Glozman:1995fu}. In the unitary chiral approaches,
$\Xi(1820)$ was even suggested to be a dynamically generated state
from the $\Delta$ decuplet-pion octet chiral
interaction~\cite{Kolomeitsev:2003kt,Sarkar:2004jh}.

Although $\Xi(1950)$ is a three-star state listed in PDG, not much
can be said about its properties~\cite{Beringer:1900zz}. According
to various phenomenological model
predictions~\cite{Chao:1980em,Matagne:2011fr,Faiman:1973br,Schat:2001xr,Goity:2002pu,Goity:2003ab},
many states with spin-parity quantum numbers, $J^P=1/2^{\pm}$,
$3/2^{\pm}$ and $5/2^{\pm}$, could be candidates for $\Xi(1950)$,
because their predicted masses are close to 1950 MeV. Alitti
\emph{et al.} suggested that the broad $\Xi$ resonance
($\Gamma=80\pm 40$ MeV) with a mass of $M=1930\pm 20$ MeV observed
by them could be classified as the pure $J^P=5/2^-$ octet
state~\cite{Alitti:1968zz}. With this assignment, the mass and total
and partial decay widths seemed to be reasonably understood.
However, the observations of Biagi \emph{et al.}~\cite{Biagi:1986vs}
do not satisfy this classification. Although they observed a $\Xi$
resonance around 1950 MeV in the $\Lambda \bar{K}$ invariant mass
distribution, its favorable spin-parity should be $5/2^+$ or its
spin should be greater than $5/2$ in the natural spin-parity series
$7/2^-$, $9/2^+$, etc~\cite{Biagi:1986vs}. In fact, there might be
more than one $\Xi$ near 1950 MeV~\cite{Beringer:1900zz}. Recently,
Valderrama, Xie and Nieves proposed the existence of three
$\Xi(1950)$ states: one of these states would be part of a
spin-parity $1/2^+$ decuplet and the other two probably would belong
to the $5/2^+$ and $5/2^-$ octets~\cite{PavonValderrama:2011gp}.

About $\Xi(2030)$, not many discussions can be found in the
literature. A moment analysis of~\cite{Hemingway:1977uw} suggested
that the spin of $\Xi(2030)$ should be $J\geq 5/2$. According to the
SU(3) symmetry of hadrons, Samios \emph{et al.} predicted that
$\Xi(2030)$ was most likely to be the partner of $N(1680)$,
$\Lambda(1820)$ and $\Sigma(1915)$ with
$J^P=5/2^+$~\cite{Samios:1974tw}. The quark model mass calculations
indicated that the second $J^P=5/2^+$ state and $J^P=7/2^+$ state
might be candidates for
$\Xi(2030)$~\cite{Chao:1980em,Pervin:2007wa}. However, the recent
strong decay analysis of~\cite{PavonValderrama:2011gp} did not
support the classification of~\cite{Samios:1974tw}.

Fortunately, the situation of the poor knowledge about these hyperon
resonances is to be changed completely with the running of the Japan
Proton Accelerator Research Complex (J-PARC) facility, where high
quality antikaon beams will be available very
soon~\cite{Nagae:2008zz}. Then, the $\Xi$ baryons can be directly
produced on the nucleons by the process $\bar{K}N\rightarrow \Xi K$
with large production cross sections. Furthermore, the investigation
of $\Xi$ baryons is also one of the major parts of the physics
programs at $\mathrm{\bar{P}ANDA}$ via the process
$\bar{p}p\rightarrow \Xi\bar{\Xi}$~\cite{Lutz:2009ff}, at Jlab via
the photo-production processes, $\gamma p\rightarrow K^+K^+\Xi^-$
and $\gamma p\rightarrow
K^+K^+\Xi^0\pi^-$~\cite{Price:2004xm,Price:2004hr,Guo:2007dw}, and
at BESIII via the process $\mathrm{Charmonium} \rightarrow
\bar{\Xi}\Xi$~\cite{Zou:2000wg,Asner:2008nq}, etc. Thus, these new
facilities provide us with good chances to study the $\Xi$ spectrum.

Stimulated by the great progress achieved in experiments, more and
more theoretical interests have begun to focus on the $\Xi$ physics.
For example, some theoretical studies of the production of $\Xi$ via
$\bar{K}N\rightarrow \Xi K$~\cite{Sharov:2011xq,Shyam:2011ys} and
$\gamma p\rightarrow K^+K^+\Xi^-$~\cite{Nakayama:2006ty,Man:2011np},
and the determination of the parity of $\Xi$ resonances in these
reactions were started recently~\cite{Nakayama:2012zp}. In this
work, we make a systematic study of the strong decays of the $\Xi$
resonances in a chiral quark model, which has been developed and
successfully used to deal with the strong decays of charmed baryons
and heavy-light
mesons~\cite{Zhong:2007gp,Liu:2012sj,Zhong:2010vq,Zhong:2009sk,Zhong:2008kd}.
Here, we mainly focus on the three-star $\Xi$ resonances
$\Xi(1690)$, $\Xi(1820)$, $\Xi(1950)$ and $\Xi(2030)$.

This work is organized as follows. In the subsequent section,
baryons in the constituent quark model are outlined. Then a brief
review of the strong decays described within a chiral quark model
approach is given in Sec.~\ref{SDOB}. The numerical results are
presented and discussed in Sec.~\ref{RA}. Finally, a summary is
given in Sec.~\ref{sum}.

\section{BARYONS IN THE QUARK MODEL}

\subsection{Spatial wave functions}

A baryon containing three quarks may be described by an oscillator
potential Hamiltonian \cite{Bhaduri} in a nonrelativistic form:
\begin{equation}\label{hm1}
\mathcal{H} =
\sum_{i=1}^3\frac{p_i^2}{2m_i}+\frac{1}{2}K\sum_{i<j}(\mathbf{r}_j-\mathbf{r}_i)^2,
\end{equation}
where $\mathbf{r}_i$ and $\mathbf{p}_i$ are the coordinate and
momentum for the $j$th quark in the baryon rest frame, respectively.
The quarks are confined in an oscillator potential with the
potential parameter $K$ independent of the flavor quantum number.
One defines the Jacobi coordinates to eliminate the center mass
(c.m.) variables:
\begin{eqnarray}
\vec{\rho}&=&\frac{1}{\sqrt{2}}(\mathbf{r}_1-\mathbf{r}_2),\label{zb1}\\
\vec{\lambda}&=&\frac{1}{\sqrt{6}}\left(\frac{m_1\mathbf{r}_1+m_2\mathbf{r}_2}{m_1+m_2}-2\mathbf{r}_3\right),\label{zb2}\\
\mathbf{R}_{c.m.}&=&\frac{m_1\mathbf{r}_1+m_2\mathbf{r}_2+m_3\mathbf{r}_3}{M}\label{zb3},
\end{eqnarray}
where $M\equiv m_1+m_2+m_3$. Using the Jacobi coordinates in Eqs.
(\ref{zb1})-(\ref{zb3}), the oscillator Hamiltonian (\ref{hm1}) is
reduced to
\begin{eqnarray} \label{hm2}
\mathcal{H}=\frac{P^2_{c.m.}}{2M
}+\frac{1}{2m_\rho}\mathbf{p}^2_\rho+\frac{1}{2m_\lambda}\mathbf{p}^2_\lambda+
\frac{3}{2}K(\rho^2+\lambda^2),
\end{eqnarray}
where
\begin{eqnarray} \label{mom}
\mathbf{p}_\rho=m_\rho\dot{\vec{\rho}},\ \
\mathbf{p}_\lambda=m_\lambda\dot{\vec{\lambda}},\ \
\mathbf{P}_{c.m.}=M \mathbf{\dot{R}}_{c.m.},
\end{eqnarray}
with
\begin{eqnarray}
m_\lambda=\frac{3(m_1+m_2)m_3}{2M},\ \
m_\rho=\frac{2m_1m_2}{m_1+m_2}.
\end{eqnarray}

For a $\Xi$ baryon, it contains two strange ($s$) quarks and a light
$u/d$ quark. According to Eqs.(\ref{zb1})-(\ref{zb3}), the
coordinates, $\mathbf{r}_1$, $\mathbf{r}_2$ and $\mathbf{r}_3$, can
be translated into functions of the Jacobi coordinates $\lambda$ and
$\rho$, which are given by
\begin{eqnarray}
\mathbf{r}_1&=&\mathbf{R}_{c.m.}+\frac{1}{\sqrt{6}}\frac{3m'}{2m+m'}\vec{\lambda}+\frac{1}{\sqrt{2}}\vec{\rho},\\
\mathbf{r}_2&=&\mathbf{R}_{c.m.}+\frac{1}{\sqrt{6}}\frac{3m'}{2m+m'}\vec{\lambda}-\frac{1}{\sqrt{2}}\vec{\rho},\\
\mathbf{r}_3&=&\mathbf{R}_{c.m.}-\sqrt{\frac{2}{3}}\frac{3m}{2m+m'}\vec{\lambda}.
\end{eqnarray}
Then, the momenta $\mathbf{p}_1$, $\mathbf{p}_2$ and $\mathbf{p}_3$
are given by
\begin{eqnarray}
\mathbf{p}_1&=&\frac{m}{M}\mathbf{P}_{c.m.}+\frac{1}{\sqrt{6}}\mathbf{p}_\lambda+\frac{1}{\sqrt{2}}\mathbf{p}_\rho,\\
\mathbf{p}_2&=&\frac{m}{M}\mathbf{P}_{c.m.}+\frac{1}{\sqrt{6}}\mathbf{p}_\lambda-\frac{1}{\sqrt{2}}\mathbf{p}_\rho,\\
\mathbf{p}_3&=&\frac{m'}{M}\mathbf{P}_{c.m.}-\sqrt{\frac{2}{3}}\mathbf{p}_\lambda,
\end{eqnarray}
where $m$ and $m'$ stand for the masses of strange and $u/d$ quarks,
respectively.

The spatial wave function is a product of the $\rho$-oscillator and
the $\lambda$-oscillator states. With the standard notation, the
principal quantum numbers of the $\rho$-oscillator and
$\lambda$-oscillator are $N_\rho=(2n_\rho+l_\rho)$ and
$N_\lambda=(2n_\lambda+l_\lambda)$, and  the energy of a state is
given by
\begin{eqnarray}
E_N&=&\left(N_\rho+\frac{3}{2}\right)\omega_\rho+\left(N_\lambda+\frac{3}{2}\right)\omega_\lambda
\ .
\end{eqnarray}
The total principal quantum number (i.e. shell number) $N$ is
defined as
\begin{eqnarray}
N=N_\rho+N_\lambda,
\end{eqnarray}
and the frequencies of the $\rho$-mode and $\lambda$-mode are
\begin{eqnarray}\label{freq}
\omega_\lambda=(3K/m_\lambda)^{1/2},\ \omega_\rho=(3K/m_\rho)^{1/2}.
\end{eqnarray}

In the constituent quark model, a useful oscillator parameter, i.e.
the potential strength is defined as
\begin{equation}
\alpha_\lambda=(m_\lambda\omega_\lambda)^{1/2}=\left(\frac{3m'}{2m+m'}\right)^{1/4}\alpha_\rho.
\end{equation}
For a $\Xi$ baryon, the constituent mass of a strange quark is close
to that of a $u/d$ quark. Thus, the relation
\begin{equation}
\alpha_\lambda\simeq\alpha_\rho\equiv \alpha_h,
\end{equation}
is a good approximation.

The total spatial wave function of the oscillator Hamiltonian can be
written as~\cite{Karl:1969iz,Zhenping:1991}
\begin{equation}
\Psi(\mathbf{r}_1,\mathbf{r}_2,\mathbf{r}_3)=\frac{e^{i\mathbf{P}_{c.m.}
\cdot\mathbf{R}_{c.m.}}}{\sqrt{(2\pi)^3}}\Psi^\sigma_{NLL_Z}(\vec{\rho},\vec{\lambda}),
\end{equation}
where $\Psi^\sigma_{NLL_Z}(\vec{\rho},\vec{\lambda})$ is the
harmonic oscillator wave function, and the $\sigma=s,\rho,\lambda,a$
stands for the representation of the $S_3$ group. The harmonic
oscillator wave function is given
by~\cite{Karl:1969iz,Zhenping:1991}
\begin{eqnarray}
\Psi^\sigma_{NLL_Z}(\vec{\rho},\vec{\lambda})=P^\sigma_{NLL_z}\frac{\alpha^3_h}
{\pi^{3/2}}e^{-\alpha^2_h(\vec{\rho}^2+\vec{\lambda}^2)/2},
\end{eqnarray}
where $P^\sigma_{NLL_z}$ is a polynomial of $\rho$ and $\lambda$.
Explicitly, the states within $N\leq2$ can be
obtained~\cite{Zhenping:1991}:
\begin{eqnarray}
   N=0,~~L=0,~~P^s_{000}&=&1,\\
   N=1,~~L=1,~~P^\rho_{111}&=&\alpha_h\rho_+,\\
               P^\lambda_{111}&=&\alpha_h\lambda_+,\\
   N=2,~~L=0,~~P^s_{200}&=& \frac{\alpha^2_h}{\sqrt{3}}(\vec{\rho}^2+\vec{\lambda}^2-3\alpha^{-2}_h),\\
             ~~P^\rho_{200}&=&\frac{\alpha^2_h}{\sqrt{3}}2\vec{\rho}\cdot\vec{\lambda },\\
             ~~P^\lambda_{200}&=& \frac{\alpha^2_h}{\sqrt{3}}(\vec{\rho}^2-\vec{\lambda}^2),\\
       ~~L=2,~~P^s_{222}&=&\frac{1}{2}\alpha^2_h(\rho^2_++\lambda^2_+),\\
             ~~P^\rho_{222}&=&\alpha^2_h\rho_+\lambda_+,\\
             ~~P^\lambda_{222}&=&\frac{1}{2}\alpha^2_h(\rho^2_+-\lambda^2_+),\\
       ~~L=1,~~P^a_{211}&=&\alpha^2_h(\rho_+\lambda_z-\lambda_+\rho_z).
\end{eqnarray}

\subsection{Flavor and spin wave functions}

In the quark model, the SU(3) flavor wave functions of the symmetry
decuplet are~\cite{Zhenping:1991}
       \begin{eqnarray}
     \phi^s =\cases{uuu,  &$\Delta^{++}$\cr
         \frac{1}{\sqrt{3}}(uud+udu+duu),  &$\Delta^+$\cr
         \frac{1}{\sqrt{3}}(uus+usu+suu), &$\Sigma^{*+}$\cr
         ddd,  &$\Delta^-$\cr
         \frac{1}{\sqrt{3}}(udd+dud+ddu),  &$\Delta^0$\cr
         \frac{1}{\sqrt{3}}(dds+dsd+sdd),  &$\Sigma^{*-}$\cr
         sss,  &$\Omega$\cr
         \frac{1}{\sqrt{3}}(uss+sus+ssu),  &$\Xi^{*0}$\cr
         \frac{1}{\sqrt{3}}(dss+sds+ssd), &$\Xi^{*-}$\cr
         \frac{1}{\sqrt{6}}(uds+sud+dsu+sdu+dus+usd),
         &$\Sigma^{*0}$}
 \end{eqnarray}
the mixed-symmetry octet wave functions are~\cite{Zhenping:1991}
    \begin{eqnarray}
     \phi^\lambda =\cases{
      \frac{1}{\sqrt{6}}(2uud-duu-udu),   &p\cr
      \frac{1}{\sqrt{6}}(dud+udd-2ddu),  &n\cr
       \frac{1}{\sqrt{6}}(2uus-suu-usu),  &$\Sigma^+$\cr
       \frac{1}{2\sqrt{3}}(sdu+sud+usd+dsu-2uds-2dus),  &$\Sigma^0$\cr
      \frac{1}{\sqrt{6}}(2dds-sdd-dsd),  &$\Sigma^-$\cr
       \frac{1}{2}(sud+usd-sdu-dsu),  &$\Lambda$\cr
       \frac{1}{\sqrt{6}}(sus+uss-2ssu),  &$\Xi^0$\cr
        \frac{1}{\sqrt{6}}(dss+sds-2ssd),  &$\Xi^-$}
     \end{eqnarray}
while the mixed-antisymmetric octet wave functions
are~\cite{Zhenping:1991}
     \begin{eqnarray}
     \phi^\rho =\cases{
     \frac{1}{\sqrt{2}}(udu-duu),   &p\cr
      \frac{1}{\sqrt{2}}(udd-dud),  &n\cr
       \frac{1}{\sqrt{2}}(usu-suu),  &$\Sigma^+$\cr
       \frac{1}{2}(sud+sdu-usd-dsu),  &$\Sigma^0$\cr
      \frac{1}{\sqrt{2}}(dsd-sdd),  &$\Sigma^-$\cr
       \frac{1}{2\sqrt{3}}(usd+sdu-sud-dsu-2dus+2uds),  &$\Lambda$\cr
       \frac{1}{\sqrt{2}}(uss-sus),  &$\Xi^0$\cr
        \frac{1}{\sqrt{2}}(dss-sds).  &$\Xi^-$}
      \end{eqnarray}
The total antisymmetric singlet flavor wave function is
  \begin{equation}
     \phi^a=\frac{1}{\sqrt{6}}(uds+dsu+sud-dus-usd-sdu).
  \end{equation}

In the quark model, the typical SU(2) spin wave functions for the
baryons can be adopted \cite{Koniuk:1979vy,Copley:1979wj}:
\begin{eqnarray}
\chi^s_{3/2}&=&\uparrow\uparrow\uparrow, \ \  \chi^s_{-3/2}=\downarrow\downarrow\downarrow, \nonumber\\
\chi^s_{1/2}&=&\frac{1}{\sqrt{3}}(\uparrow\uparrow\downarrow+\uparrow\downarrow\uparrow+
\downarrow\uparrow\uparrow),\nonumber\\
\chi^s_{-1/2}&=&\frac{1}{\sqrt{3}}(\uparrow\downarrow\downarrow+\downarrow\downarrow
\uparrow+\downarrow\uparrow\downarrow),
\end{eqnarray}
for the spin-3/2 states with symmetric spin wave functions,
\begin{eqnarray}
\chi^\rho_{1/2}&=&\frac{1}{\sqrt{2}}(\uparrow\downarrow\uparrow-\downarrow\uparrow\uparrow),\nonumber\\
\chi^\rho_{-1/2}&=&\frac{1}{\sqrt{2}}(\uparrow\downarrow\downarrow-\downarrow\uparrow\downarrow),
\end{eqnarray}
for the spin-1/2 states with mixed antisymmetric spin wave
functions, and
\begin{eqnarray}
\chi^\lambda_{1/2}&=&-\frac{1}{\sqrt{6}}(\uparrow\downarrow\uparrow+\downarrow\uparrow
\uparrow-2\uparrow\uparrow\downarrow),\nonumber\\
\chi^\lambda_{-1/2}&=&+\frac{1}{\sqrt{6}}(\uparrow\downarrow\downarrow+\downarrow
\uparrow\downarrow-2\downarrow\downarrow\uparrow),
\end{eqnarray}
for the spin-1/2 states with mixed symmetric spin wave functions.

The flavor-spin wave functions are representations of SU(6), which
are denoted by $|N_6,^{2S+1}N_3\rangle$, where $N_6$ ($N_3$)
represents the SU(6) (SU(3)) representation and $S$ stands for the
total spin of the wave function. The symmetric $\mathbf{56}$
representation wave functions are~\cite{Zhenping:1991}
\begin{eqnarray}
|56,~^{2}8\rangle^s&=&\frac{1}{\sqrt{2}}(\phi^\rho\chi^\rho+\phi^\lambda\chi^\lambda),\\
|56,^{4}10\rangle^s&=&\phi^s\chi^s,
\end{eqnarray}
the antisymmetric $\mathbf{70}$ representation wave functions
are~\cite{Zhenping:1991}
\begin{eqnarray}
|70,~^{2}8\rangle^\rho&=&\frac{1}{\sqrt{2}}(\phi^\rho\chi^\lambda+\phi^\lambda\chi^\rho),\\
|70,~^{4}8\rangle^\rho&=&\phi^\rho\chi^s,\\
|70,^{2}10\rangle^\rho&=&\phi^s\chi^\rho,\\
|70,~^{2}1\rangle^\rho&=&\phi^\alpha\chi^\lambda,
 \end{eqnarray}
while the symmetric $\mathbf{70}$ representation wave functions
are~\cite{Zhenping:1991}
\begin{eqnarray}
|70,~^{2}8\rangle^\lambda&=&\frac{1}{\sqrt{2}}(\phi^\rho\chi^\rho-\phi^\lambda\chi^\lambda),\\
|70,~^{4}8\rangle^\lambda&=&\phi^\lambda\chi^s,\\
|70,^{2}10\rangle^\lambda&=&\phi^s\chi^\lambda,\\
|70,~^{2}1\rangle^\lambda&=&\phi^\alpha\chi^\rho,
\end{eqnarray}
and the antisymmetric $\mathbf{20}$ representation wave functions
are~\cite{Zhenping:1991}
\begin{eqnarray}
|20,^{2}8\rangle^a&=&\frac{1}{\sqrt{2}}(\phi^\rho\chi^\lambda-\phi^\lambda\chi^\rho),\\
|20,^{4}1\rangle^a&=&\phi^a\chi^s.
\end{eqnarray}

\subsection{Total wave functions}

The total spin-flavor-space wave functions are the SU(6)$\otimes$
O(3) representations,  which are denoted by
\begin{eqnarray}
|SU(6)\otimes O(3)\rangle=|N_{6},^{2S+1}N_{3},N, L,J^P\rangle.
\end{eqnarray}
The spin-flavor-space wave functions have to be
permutation-symmetric because the baryon color wave function is
totally anti-symmetric. The spin-flavor-space wave functions up to
$N=2$ shell are listed in Tab.~\ref{wfL}. So far no experimental
evidence for the existence of the $\mathbf{20}$ SU(6) representation
has been found, thus, we do not consider these excitations in this
work.

\begin{table}[ht]
\caption{The spin-flavor-space wave functions of different light
baryons, denoted by $|\mathbf{N}_{6},^{2S+1}\mathbf{N}_{3},N,
L,J^P\rangle$. The Clebsch-Gordan series for the spin and
angular-momentum addition $|J,J_z\rangle$= $\sum_{L_z+S_z=J_z}
\langle LL_z,SS_z|JJ_z \rangle \Psi^{\sigma }_{NLL_z} \chi_{S_z}$
has been omitted. } \label{wfL}
\begin{tabular}{|c|c c|c|c|c|c }\hline\hline
state & N  &\textbf{J} \ L \ \ $S$\ \   \ \ &\ \ wave function   \\
\hline
$|56,^{2}8,0,0,\frac{1}{2}^+ \rangle$& 0 & \ $\frac{1}{2}$ \ \ 0 \ \ $\frac{1}{2}$ \ \     & $|56,^28\rangle\Psi^s_{000}$      \\
\hline
$|56,^{4}10,0,0,\frac{3}{2}^+ \rangle$&0 & \ $\frac{3}{2}$ \ \ 0 \ \ $\frac{3}{2}$ \ \     &  $ |56,^410\rangle\Psi^s_{000} $                     \\
\hline
$|56,^{2}8,2,0,\frac{1}{2}^+ \rangle$&2& \ $\frac{1}{2}$ \ \ 0 \ \ $\frac{1}{2}$ \ \  &    $|56,^28\rangle\Psi^s_{200}$       \\
\hline
$|56,^{4}{10},2,0,\frac{3}{2}^+ \rangle$&2& \ $\frac{3}{2}$ \ \ 0 \ \ $\frac{3}{2}$ \ \  &  $  |56,^410\rangle\Psi^s_{200} $           \\
\hline
$|56,^{2}{8},2,2,\frac{3}{2}^+ \rangle$&2& \ $\frac{5}{2}$ \ \ 2 \ \ $\frac{1}{2}$ \ \  &  $|56,^28\rangle\Psi^s_{22L_z}$       \\
$|56,^{2}{8},2,2,\frac{5}{2}^+ \rangle$&2& \ $\frac{5}{2}$ \ \ 2 \ \ $\frac{1}{2}$ \ \  &        \\
\hline
$|56,^{4}{10},2,2,\frac{1}{2}^+ \rangle$&2& \ $\frac{7}{2}$ \ \ 2 \ \ $\frac{3}{2}$ \ \  &         \\
$|56,^{4}{10},2,2,\frac{3}{2}^+ \rangle$&2& \ $\frac{7}{2}$ \ \ 2 \ \ $\frac{3}{2}$ \ \   &   $ |56,^410\rangle\Psi^s_{22L_z} $\\
$|56,^{4}{10},2,2,\frac{5}{2}^+ \rangle$&2& \ $\frac{7}{2}$ \ \ 2 \ \ $\frac{3}{2}$ \ \  &   \\
$|56,^{4}{10},2,2,\frac{7}{2}^+ \rangle$&2& \ $\frac{7}{2}$ \ \ 2 \ \ $\frac{3}{2}$ \ \  &   \\
\hline
$|70,^{2}{8},1,1,\frac{1}{2}^- \rangle$&1& \  $\frac{3}{2}$ \ \  1 \ \ $\frac{1}{2}$ \ \    & $|70,^{2}8\rangle^\rho\Psi^\rho_{11L_z}+|70,^{2}8\rangle^\lambda\Psi^\lambda_{11L_z}$     \\
$|70,^{2}{8},1,1,\frac{3}{2}^- \rangle$&1& \  $\frac{3}{2}$ \ \  1 \ \ $\frac{1}{2}$ \ \  &\\
\hline
$|70,^{4}{8},1,1,\frac{1}{2}^- \rangle$&1& \  $\frac{5}{2}$ \ \  1 \ \ $\frac{3}{2}$ \ \  &\\
$|70,^{4}{8},1,1,\frac{3}{2}^- \rangle$&1& \  $\frac{5}{2}$ \ \  1 \ \ $\frac{3}{2}$ \ \  &$|70,^{4}8\rangle^\rho\Psi^\rho_{11L_z}+|70,^{4}8\rangle^\lambda\Psi^\lambda_{11L_z}$\\
$|70,^{4}{8},1,1,\frac{5}{2}^- \rangle$&1& \  $\frac{5}{2}$ \ \  1 \ \ $\frac{3}{2}$ \ \  &\\
\hline
$|70,^{2}{10},1,1,\frac{1}{2}^- \rangle$&1& \ $\frac{3}{2}$ \ \ 1 \ \ $\frac{1}{2}$ \ \      &  $|70,^{2}{10}\rangle^\rho\Psi^\rho_{11L_z}+|70,^{2}{10}\rangle^\lambda\Psi^\lambda_{11L_z}$     \\
$|70,^{2}{10},1,1,\frac{3}{2}^- \rangle$&1& \ $\frac{3}{2}$ \ \ 1 \ \ $\frac{1}{2}$ \ \      &          \\
\hline
$|70,^{2}{8},2,0,\frac{1}{2}^+ \rangle$&2 & \ $\frac{1}{2}$ \ \ 0 \ \ $\frac{1}{2}$ \ \   &$  |70,^{2}{8}\rangle^\rho\Psi^\rho_{200}+|70,^{2}{8}\rangle^\lambda\Psi^\lambda_{200}$     \\
\hline
$|70,^{4}{8},2,0,\frac{3}{2}^+ \rangle$&2 & \ $\frac{3}{2}$ \ \ 0 \ \ $\frac{3}{2}$ \ \   &$  |70,^{4}{8}\rangle^\rho\Psi^\rho_{200}+|70,^{4}{8}\rangle^\lambda\Psi^\lambda_{200}$     \\
\hline
$|70,^{2}{10},2,0,\frac{1}{2}^+ \rangle$&2 & \ $\frac{1}{2}$ \ \ 0 \ \ $\frac{1}{2}$ \ \   &$  |70,^{2}{10}\rangle^\rho\Psi^\rho_{200}+|70,^{2}{10}\rangle^\lambda\Psi^\lambda_{200}$     \\
\hline
$|70,^{2}{8},2,2,\frac{3}{2}^+ \rangle$&2 & \ $\frac{5}{2}$ \ \ 2 \ \ $\frac{1}{2}$ \ \      &$  |70,^{2}{8}\rangle^\rho\Psi^\rho_{22L_z}+|70,^{2}{8}\rangle^\lambda\Psi^\lambda_{22L_z}$      \\
$|70,^{2}{8},2,2,\frac{5}{2}^+ \rangle$&2 & \ $\frac{5}{2}$ \ \ 2 \ \ $\frac{1}{2}$ \ \     &          \\
\hline
$|70,^{4}{8},2,2,\frac{1}{2}^+ \rangle$&2 & \ $\frac{7}{2}$ \ \ 2 \ \ $\frac{3}{2}$ \ \      &       \\
$|70,^{4}{8},2,2,\frac{3}{2}^+ \rangle$&2 & \ $\frac{7}{2}$ \ \ 2 \ \ $\frac{3}{2}$ \ \       & $ |70,^{4}{8}\rangle^\rho\Psi^\rho_{22L_z}+|70,^{4}{8}\rangle^\lambda\Psi^\lambda_{22L_z}$      \\
$|70,^{4}{8},2,2,\frac{5}{2}^+ \rangle$&2 & \ $\frac{7}{2}$ \ \ 2 \ \ $\frac{3}{2}$ \ \      &       \\
$|70,^{4}{8},2,2,\frac{7}{2}^+ \rangle$&2 & \ $\frac{7}{2}$ \ \ 2 \ \ $\frac{3}{2}$ \ \      &       \\
\hline
$|70,^{2}{10},2,2,\frac{3}{2}^+ \rangle$&2 & \ $\frac{5}{2}$ \ \ 2 \ \ $\frac{1}{2}$ \ \     &  $ |70,^{2}{10}\rangle^\rho\Psi^\rho_{22L_z}+|70,^{2}{10}\rangle^\lambda\Psi^\lambda_{22L_z}$      \\
$|70,^{2}{10},2,2,\frac{5}{2}^+ \rangle$&2 & \ $\frac{5}{2}$ \ \ 2 \ \ $\frac{1}{2}$ \ \       &      \\
\hline
\end{tabular}
\end{table}

\section{strong decay of a baryon in the chiral quark model}\label{SDOB}

In the chiral quark model, the effective low energy quark-meson
pseudoscalar coupling at tree level is given
by~\cite{Li:1995si,Zhong:2007fx,Li:1997gda,qk3}
\begin{eqnarray}\label{coup}
H_m=\sum_j
\frac{1}{f_m}\bar{\psi}_j\gamma^{j}_{\mu}\gamma^{j}_{5}\psi_j\vec{\tau}\cdot
\partial^{\mu}\vec{\phi}_m,
\end{eqnarray}
where $\psi_j$ represents the $j$th quark field in a baryon, and
$f_m$ is the meson's decay constant. The pseudoscalar meson octet
$\phi_m$ is expressed as
\begin{eqnarray}
\phi_m=\pmatrix{
 \frac{1}{\sqrt{2}}\pi^0+\frac{1}{\sqrt{6}}\eta & \pi^+ & K^+ \cr
 \pi^- & -\frac{1}{\sqrt{2}}\pi^0+\frac{1}{\sqrt{6}}\eta & K^0 \cr
 K^- & \bar{K}^0 & -\sqrt{\frac{2}{3}}\eta}.
\end{eqnarray}
To match the nonrelativistic harmonic oscillator spatial wave
function $^N\Psi_{LL_z}$ in the quark model, we adopt the
nonrelativistic form of Eq. (\ref{coup}) in the calculations, which
is given by \cite{Li:1995si,Zhong:2007fx,Li:1997gda,qk3}
\begin{eqnarray}\label{non-relativistic-expans}
H^{nr}_{m}&\simeq&\sum_{j=1}^{3}\Big\{\frac{\omega_m}{E_f+M_f}\vsig_j\cdot
\textbf{P}_f+ \frac{\omega_m}{E_i+M_i}\vsig_j \cdot
\textbf{P}_i \nonumber\\
&&-\vsig_j \cdot \textbf{q} +\frac{\omega_m}{2\mu_q}\vsig_j\cdot
\textbf{p}'_j\Big\}I_j \varphi_m,
\end{eqnarray}
where $\vsig_j$ is the Pauli spin operator on the the $j$th quark of
the baryon, and $\mu_q$ is a reduced mass given by
$1/\mu_q=1/m_j+1/m_j'$ with $m_j$ and $m_j'$ for the masses of the
$j$th quark in the initial and final baryons, respectively. For
emitting a meson, we have $\varphi_m=e^{-i\textbf{q}\cdot
\textbf{r}_j}$, and for absorbing a meson we have
$\varphi_m=e^{i\textbf{q}\cdot \textbf{r}_j}$. In the above
nonrelativistic expansion, $\textbf{p}'_j$ ($ =\textbf{p}_j-m_j/M
\textbf{P}_{c.m.}$) is the internal momentum operator for the $j$th
quark in the baryon rest frame. $\omega_m$ and $\textbf{q}$ are the
energy and three-vector momentum of the meson, respectively.
$\textbf{P}_i$ and $ \textbf{P}_f$ stand for the momenta of the
initial final baryons, respectively. The isospin operator $I_j$ in
Eq. (\ref{non-relativistic-expans}) is expressed as
\begin{eqnarray}
I_j=\cases{ a^{\dagger}_j(u)a_j(s) & for $K^+$, \cr
a^{\dagger}_j(s)a_j(u) & for $K^-$,\cr a^{\dagger}_j(d)a_j(s) & for
$K^0$, \cr a^{\dagger}_j(s)a_j(d) & for $\bar{K^0}$,\cr
a^{\dagger}_j(u)a_j(d) & for $\pi^-$,\cr a^{\dagger}_j(d)a_j(u)  &
for $\pi^+$,\cr
\frac{1}{\sqrt{2}}[a^{\dagger}_j(u)a_j(u)-a^{\dagger}_j(d)a_j(d)] &
for $\pi^0$, \cr
\frac{1}{\sqrt{2}}[a^{\dagger}_j(u)a_j(u)+a^{\dagger}_j(d)a_j(d)]\cos\phi_P
\cr - a^{\dagger}_j(s)a_j(s)\sin\phi_P & for $\eta$,}
\end{eqnarray}
where $a^{\dagger}_j(u,d,s)$ and $a_j(u,d,s)$ are the creation and
annihilation operators for the $u$, $d$ and $s$ quarks, and $\phi_P$
is the mixing angle of $\eta$ meson in the flavor
basis~\cite{Beringer:1900zz,Zhong:2011ht}.

In the calculations, we select the initial-baryon-rest system for
the decay processes. The energies and momenta of the initial baryons
are denoted by $(E_i, \textbf{P}_i$), while those of the final state
mesons and baryons are denoted by $(\omega_f, \textbf{q})$ and
$(E_f, \textbf{P}_f)$, respectively. Note that $\textbf{P}_i=0$ and
$\textbf{P}_f=-\textbf{q}$. For a light pseudoscalar meson emission
in the strong decay process of a baryon, $\mathcal{B}\to\mathcal{B}'
\mathbb{M}(\mathbf{q})$, the partial decay amplitudes can be worked
out according to
\begin{eqnarray}
\mathcal{M}[\mathcal{B}\to   \mathcal{B}' \mathbb{M}(\mathbf{q})] =
3\left\langle \mathcal{B}'\left|\left\{G\vsig_3\cdot \textbf{q}
+h\vsig_3\cdot \textbf{p}'_3\right\}I_3 e^{-i\textbf{q}\cdot
\textbf{r}_3}\right|\mathcal{B}\right\rangle,
\end{eqnarray}
with
\begin{eqnarray}\label{ccpk}
G\equiv -\frac{\omega_m}{E_f+M_f}-1,\ \ h\equiv
\frac{\omega_m}{2\mu_q},
\end{eqnarray}
where $\mathcal{B}$ and $\mathcal{B}'$ stand for the initial and
final baryon wave functions listed in Tab.~\ref{wfL}.

With the derived decay amplitudes, one can calculate the width
by~\cite{Zhong:2007gp}
\begin{equation}\label{dww}
\Gamma=\left(\frac{\delta}{f_m}\right)^2\frac{(E_f +M_f)|q|}{4\pi
M_i}\frac{1}{2J_i+1}\sum_{J_{iz}J_{fz}}^{}|\mathcal{M}_{J_{iz},J_{fz}}|,
\end{equation}
where the $J_{iz}$ and $J_{fz}$ stand for the the total angular
momenta of the initial and final baryons, respectively. $\delta$ as
a global parameter accounts for the strength of the quark-meson
couplings.

In the calculation, the standard quark model parameters are adopted.
Namely, we set $m_u=m_d=350$ MeV, $m_s=450$ MeV, for the constituent
quark masses. The harmonic oscillator parameter $\alpha_h$ in the
wave function $^N\Psi_{LL_z}$ is taken as $\alpha_h=0.40$ GeV. The
decay constants for $\pi$, $K$ and $\eta$ mesons are taken as
$f_{\pi}=132$ MeV, $f_K=f_{\eta}=160$ MeV, respectively. The masses
of the mesons and baryons used in the calculations are adopted from
the Particle Data Group~\cite{Beringer:1900zz}.

\section{results AND ANALYSIS}\label{RA}

\begin{table}[ht]
\caption{The partial and total decay widths (MeV) of the
well-established baryons $\Xi(1530)$, $\Sigma(1385)$ and
$\Delta(1232)$, which correspond to the same representation
$|56,^{4}10,0,0,\frac{3}{2}^+\rangle$. The experimental data are
obtained from PDG.} \label{Xi1530}
\begin{tabular}{|c|c|c|c|c|c|c|c|c|c|c }\hline\hline
state &channel~&$\Gamma^{th}_{i} $&~$\Gamma^{th}_{total}$ &$\Gamma^{exp}_{total}$\\
\hline
$\Xi(1530)^0$                     &$\Xi^-\pi^+$ &  5.6              &   9.1(input)     &  $9.1\pm 0.5$\\
                                  &$\Xi^0\pi^0$ &   3.5               &        &          \\
\hline
$\Xi(1530)^-$                     &$\Xi^0\pi^-$ &   7.1               &  10.3      & $9.9^{+ 1.7}_{-1.9}$ \\
                                  &$\Xi^-\pi^0$ &     3.2             &        &   \\
\hline
$\Sigma(1385)^+$                  &$\Sigma^+\pi^0$ &  1.9           & 27.1  &$36\pm 0.7$ \\
                                  &$\Sigma^0\pi^+$ &    1.5          &    & \\
                                  &$\Lambda^0\pi^+$  &  23.7        & &  \\
\hline
$\Sigma(1385)^0$                  &$\Sigma^+\pi^-$ &   1.8           &  27.9 & $36\pm 5$ \\
                                  &$\Sigma^-\pi^+$ &   1.3        &    & \\
                                  &$\Lambda^0\pi^0$  &  24.8          & &  \\
\hline
$\Sigma(1385)^-$                  &$\Sigma^-\pi^0$ &  1.7           &  28.7  &$39.4\pm 2.1$ \\
                                  &$\Sigma^0\pi^-$ &    1.8          &    & \\
                                  &$\Lambda^0\pi^-$  &  25.2         & &  \\
\hline
$\Delta(1232)^{++}$               &$p\pi^+$ &  63.9          & 63.9 & $114\sim 120$ \\
\hline
$\Delta(1232)^+$                  &$p\pi^0$ &   43.7           & 64.7 &  $114\sim 120$\\
                                  &$n\pi^+$ &     21.0         &    & \\
\hline
$\Delta(1232)^0$                  &$p\pi^-$ &   21.3         & 64.3&  $114\sim 120$\\
                                  &$n\pi^0$ &     43.0       &    & \\
\hline
\end{tabular}
\end{table}

\subsection{$\Xi$(1530)}

$\Xi(1530)$ is the only $\Xi$ resonance whose properties are all
reasonably well known.  According to the classification of the quark
model, it is assigned to the $|56,^{4}10,0,0,\frac{3}{2}^+ \rangle$
representation.  In the present work, the measured width for
$\Xi(1530)^0\to \Xi \pi$ as an input ( i.e. $\Gamma=9.1$
MeV~\cite{Beringer:1900zz}) to determine parameter $\delta$ in
Eq.(\ref{dww}), which gives
\begin{eqnarray}
\delta=0.576.
\end{eqnarray}

With this determined parameter, we can predict the other $\Xi$
resonance strong decays. First, the strong decays of $\Xi(1530)^-$
are calculated. The results have been listed in Tab.~\ref{Xi1530},
where we find that our predictions are in agreement with the data.
Furthermore, we have noted that $\Sigma(1385)$ and $\Delta(1232)$
also correspond to the $|56,^{4}10,0,0,\frac{3}{2}^+ \rangle$
representation. They are the SU(3)-flavor counterparts of
$\Xi(1530)$. Applying the determined value for $\delta$, we
calculate the strong decays of the $\Sigma(1385)$ and $\Delta(1232)$
resonances, the results are listed in Tab.~\ref{Xi1530} as well.

It is seen that the predicted widths for $\Sigma(1385)$ are
compatible with the experimental data, and the partial decay width
ratio
\begin{equation}
\frac{\Gamma[\Sigma(1385)\rightarrow
\Sigma\pi]}{\Gamma[\Sigma(1385)\rightarrow \Lambda\pi]} \simeq
0.134\pm 0.09,
\end{equation}
are in good agreement with the PDG value~\cite{Beringer:1900zz}.

The predicted widths for $\Delta(1232)$ are underestimated by about
a factor of 2 compared with the data. In fact, there are systemic
underestimations of the decay widths for the $\Delta$ and nucleon
excitations as well. The reason is that the parameters used in this
work are only fine tuned for the $\Xi$ resonances to make more
reliable predictions for the their decay properties. For simplicity,
we do not carry out a global fit to strange and non-strange baryons.
We have noted that the global parameter $\delta=0.576$ used for the
$\Xi$ resonances in this work is slightly smaller than that used for
the $\Delta$ and nucleon resonances in~\cite{An:2011sb}.

As a whole, the predicted strong decay properties for $\Xi(1530)^-$,
$\Sigma(1385)$ are compatible with the experimental data. If a
global fit to strange and non-strange baryons is carried out, the
predicted strong decay width of $\Delta(1232)$ can be more close to
the data. Thus, the SU(3)-flavor symmetry approximately holds in
these ground decuplet baryons.

\subsection{$\Xi(1690)$}\label{mix-16}

$\Xi(1690)$ is a three-star state listed in PDG. Its decay width
might be less than $30$ MeV~\cite{Beringer:1900zz}. Three decay
modes $\Lambda \bar{K}$, $\Sigma \bar{K}$ and $\Xi\pi$ were observed
in experiments. Recently, \emph{BABAR} Collaboration found some
evidence that the $\Xi^0(1690)$ has $J^P=1/2^-$ in the study of
$\Lambda_c^+\rightarrow \Xi^- \pi^+K^+$~\cite{Aubert:2008ty}.

If $\Xi(1690)$ is a $J^P=1/2^-$ resonance, it should correspond to
one of the first orbital excitation states:
$|70,^28,1,1,\frac{1}{2}^-\rangle$,
$|70,^48,1,1,\frac{1}{2}^-\rangle$ and
$|70,^210,1,1,\frac{1}{2}^-\rangle$ or a mixed state between them.
Considering $\Xi(1690)$ as the $|70,^28,1,1, \frac{1}{2}^-\rangle$,
$|70,^48,1,1,\frac{1}{2}^-\rangle$ and
$|70,^210,1,1,\frac{1}{2}^-\rangle$, respectively, we calculate
their decay properties, which are listed in Tab.~\ref{w1690}.
Comparing the predictions with the measurements, we find that if
$\Xi(1690)$ is assigned to $|70,^28,1,1,\frac{1}{2}^-\rangle$, the
calculated decay width
\begin{eqnarray}
\Gamma\approx 48~\mathrm{MeV},
\end{eqnarray}
and partial decay width ratios
\begin{eqnarray}
\frac{\Gamma(\Xi\pi)}{\Gamma(\Sigma\bar{K})}\simeq
0.2,~\frac{\Gamma(\Sigma\bar{K})}{\Gamma(\Lambda\bar{K})}\approx1.0,
~\frac{\Gamma(\Xi(1530)\pi)}{\Gamma(\Sigma\bar{K})}\approx0.0002,
\end{eqnarray}
are roughly in agreement with the measurements. Neither
$|70,^48,1,1,\frac{1}{2}^-\rangle$ nor
$|70,^210,1,1,\frac{1}{2}^-\rangle$ could be considered as an
assignment to $\Xi(1690)$ because their partial decay width ratios
disagree with the observations.

\begin{table}[ht]
\caption{The total and partial decay widths (MeV) of $\Xi(1690)$
with different $J^P=1/2^-$ assignments.} \label{w1690}
\begin{tabular}{|c|c|c|c|c|c|c|c|c|c|c }\hline\hline
assignment &~~$~~\Xi\pi$~~&~~$\Sigma \bar{K} $~~&~~$\Lambda \bar{K} $ &$\Xi(1530)\pi$~~&~~$\Gamma_{total}$\\
\hline
$|70,^{2}8,1,1,\frac{1}{2}^-\rangle$                  &3.69  &  22.30          & 22.15  & 0.005 &48.14 \\
\hline
$|70,^{4}8,1,1,\frac{1}{2}^-\rangle$                   &59.05  & 5.58           &22.15 & 0.001 &86.78 \\
\hline
$|70,^{2}10,1,1,\frac{1}{2}^-\rangle$                  &3.69 &  1.39          &5.54  &0.005 &10.63 \\
\hline
\end{tabular}
\end{table}

As we know, configuration mixing between several states with the
same $J^P$ often occurs via some interactions. Thus, $\Xi(1690)$
might be a mixed state between $|70,^28,1,1,\frac{1}{2}^-\rangle$,
$|70,^48,1,1,\frac{1}{2}^-\rangle$ and
$|70,^210,1,1,\frac{1}{2}^-\rangle$. Now, we consider the physical
states with $J^P=\frac{1}{2}^-$ as mixed states between
$|70,^28\rangle$, $|70,^48\rangle$ and $|70,^210\rangle$. According
to the standard CKM matrix  method, the physical states can be
express as
\begin{equation}\label{mix1690}
\left(\begin{array}{c}|\Xi \frac{1}{2}^-\rangle_1 \cr |\Xi
\frac{1}{2}^-\rangle_2 \cr |\Xi \frac{1}{2}^-\rangle_3
\end{array}\right) =U \left(\begin{array}{c} |70,^28\rangle \cr
|70,^48\rangle \cr |70,^210\rangle  \end{array}\right),
\end{equation}
with
\begin{equation}
U=\left(\begin{array}{ccc} c_{12}c_{13} & s_{12}c_{13} & s_{13} \cr
-s_{12}c_{23}-c_{12}s_{23}s_{13} & c_{12}c_{23}-s_{12}s_{23}s_{13} &
s_{23}c_{13} \cr s_{12}s_{23}-c_{12}c_{23}s_{13} &
-c_{12}s_{23}-s_{12}c_{23}s_{13} & c_{23}c_{13}
\end{array}\right),
\end{equation}
where $c_{ij}\equiv\cos\theta_{ij}$ and
$s_{ij}\equiv\sin\theta_{ij}$ with $\theta_{ij}$ the mixing angles
to be determined by the experimental data.

In present work, we take $\Xi(1690)$ as $|\Xi
\frac{1}{2}^-\rangle_2$. By fitting the experimental data of DIONISI
78~\cite{Dionisi:1978tg} (see Tab.~\ref{w1690mix}), we have obtained
$\theta_{12}\simeq5^0$, $\theta_{13}\simeq105^0$ and
$\theta_{23}\simeq95^0$. The theoretical predictions compared with
the data were listed in Tab.~\ref{w1690mix}. From the table, we find
that the decay properties of $\Xi(1690)$ could be well described
with these determined mixing angles. Thus, $\Xi(1690)$ could be a
mixed state.

\begin{table}[ht]
\caption{The predicted total and partial decay widths (MeV) and
partial decay width ratios (MeV) of $\Xi(1690)\equiv\Psi_2$ compared
with the experiment data of DIONISI 78~\cite{Dionisi:1978tg}. The
mixing angles are $\theta_{12}=5^0$, $\theta_{13}=75^0$ and
$\theta_{23}=95^0$. } \label{w1690mix}
\begin{tabular}{|c|c|c|c|c|c|c|c|c|c|c }\hline\hline
channel~&~$\Gamma^{th}_i$~&~~$\Gamma^{th}_{total}$ ~&~$\Gamma^{exp}_{total}$~& channel ratio  & $R_{th}$&$R_{exp}$ \\
\hline
$\Xi\pi$ &1.0&37 &$44^{+23}_{-23}$ &$\Gamma(\Sigma \bar{K})/\Gamma(\Lambda \bar{K})$ &2.88&$2.7^{+0.9}_{-0.9}$\\
$\Sigma \bar{K}$   &27.0 &&&$\Gamma(\Xi\pi)/\Gamma(\Sigma \bar{K})$ &0.04 &$<0.09$\\
$\Lambda \bar{K}$ &9.4 &&&$\Gamma(\Xi(1530)\pi)/\Gamma(\Sigma \bar{K})$ &$\sim10^{-4}$ &$<0.06$\\
$\Xi(1530)\pi$&0.01&&&&&\\
\hline \hline
\end{tabular}
\end{table}

With these determined mixing angles, Eq.(\ref{mix1690}) can be
explicitly written as
\begin{equation}\label{b1690}
 \left(\begin{array}{c}|\Xi \frac{1}{2}^-\rangle_1 \cr
|\Xi(1690)\frac{1}{2}^-\rangle \cr |\Xi \frac{1}{2}^-\rangle_3
\end{array}\right)=\left(\begin{array}{ccc} -0.26& -0.02& 0.97\cr
 -0.95 & -0.17 & -0.26 \cr
 0.17 &-0.98 & 0.02
\end{array}\right)\left(\begin{array}{c} |70,^28\rangle \cr |70,^48\rangle \cr |70,^210\rangle
\end{array}\right),
\end{equation}
From the above equation, it is obviously seen that the main
component of the physical state $\Xi(1690)$ is
$|70,^28,1,1,1/2^-\rangle$ ($\sim 90\%$), which slightly mixes with
$|70,^210,1,1,1/2^-\rangle$ ($\sim 7\%$) and
$|70,^48,1,1,1/2^-\rangle$ ($\sim 3\%$).

In Ref.~\cite{Aubert:2006ux}, the \emph{BABAR} Collaboration studied
the spin of $\Xi(1690)$ from $\Lambda_c^+\rightarrow \Lambda
K^+\bar{K}^0$. They found that the spin of $\Xi(1690)$ is consistent
with the value $J=1/2$. However, they could not determine the parity
of $\Xi(1690)$. According to the mass predictions of quark model,
the mass of the first radial ($2S$) excitation of $\Xi$ with
$J^P=\frac{1}{2}^+$ might be close to $1690$
MeV~\cite{Chao:1980em,Melde:2008yr}. Thus, we should consider the
possibility of $\Xi(1690)$ as the positive parity radial ($2S$)
excitations with $J^P=\frac{1}{2}^+$. In the constituent quark
model, there are three radial ($2S$) excitations with
$J^P=\frac{1}{2}^+$: $|56,^28,2,0,\frac{1}{2}^+\rangle$,
$|70,^28,2,0,\frac{1}{2}^+\rangle$ and
$|70,^210,2,0,\frac{1}{2}^+\rangle$. In our calculation, those
states' total decay widths are too small to compare with the
experimental data. The predicted partial decay width ratios are
incompatible with the observations as well. Thus, these states with
$J^P=\frac{1}{2}^+$ as assignments to $\Xi(1690)$ should be
excluded. Furthermore, we also study the strong decay properties of
the other first orbital excitation states with $J^P=3/2^-$ and
$J^P=5/2^-$. Their decay properties are very different from those of
$\Xi(1690)$. For simplicity, the calculated results are not shown in
this work.

In a word, $\Xi(1690)$ is most likely to be the first orbital
excitation of $\Xi$ with $J^P=1/2^-$. There might exist
configuration mixing in $\Xi(1690)$. The main component of
$\Xi(1690)$ is $|70,^28,1,1,1/2^-\rangle$, which slightly mixes with
$|70,^210,1,1,1/2^-\rangle$ and $|70,^48,1,1,1/2^-\rangle$. Our
predictions are consistent with the experimental observations. The
recent mass calculations of a constituent quark model support the
classification of $\Xi(1690)$ as a $J^P =\frac{1}{2}^-$ octet
resonance~\cite{Pervin:2007wa}. The calculations of a Skyrme
model~\cite{Oh:2007cr} and unitary chiral
approaches~\cite{Kolomeitsev:2003kt,Sarkar:2004jh} also indicated
that $\Xi(1690)$ has $J^P =\frac{1}{2}^-$.

\begin{figure}[ht]
\centering \epsfxsize=8.5cm \epsfbox{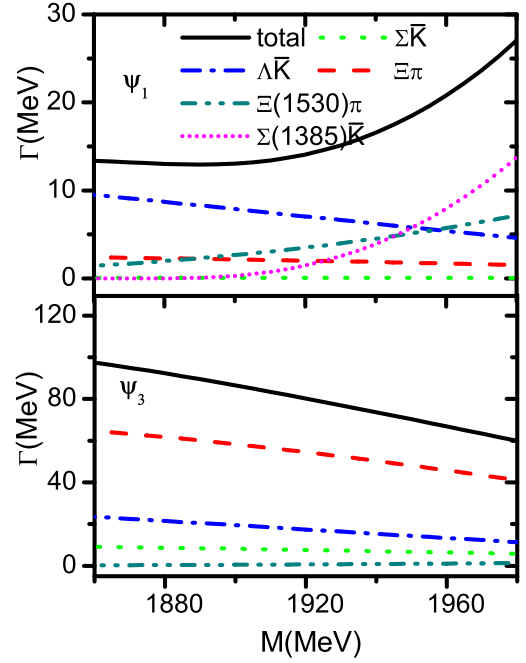} \caption{The decay
properties of $|\Xi \frac{1}{2}^-\rangle_1$ and $|\Xi
\frac{1}{2}^-\rangle_3 $, respectively, where $\Psi_1\equiv|\Xi
\frac{1}{2}^-\rangle_1 $ and $\Psi_3\equiv|\Xi
\frac{1}{2}^-\rangle_3$. }\label{fig-1690}
\end{figure}

Since $\Xi(1690)$ corresponds to the physical state $|\Xi
\frac{1}{2}^-\rangle_2$ in Eq.(\ref{b1690}), as the counterparts of
$\Xi(1690)$, the other two physical states, $|\Xi
\frac{1}{2}^-\rangle_1
$=-0.26$|70,^28\rangle-0.02|70,^48\rangle+0.97|70,^210\rangle$ and
$|\Xi \frac{1}{2}^-\rangle_3
$=0.17$|70,^28\rangle-0.98|70,^48\rangle+0.02|70,^210\rangle$, might
be observed in experiments. It is easily seen that the main
component of $|\Xi \frac{1}{2}^-\rangle_1$  is
$|70,^210,1,1,1/2^-\rangle$ ($\sim 94\%$), while $|\Xi
\frac{1}{2}^-\rangle_3 $ is dominated by $|70,^48,1,1,1/2^-\rangle$
($\sim 96\%$). According to the analysis in large $N_c$
QCD~\cite{Schat:2001xr,Goity:2002pu} and quark
model~\cite{Isgur:1978xj}, the masses of $|\Xi
\frac{1}{2}^-\rangle_1$ and $|\Xi \frac{1}{2}^-\rangle_3 $ are
around 1920 MeV.  The strong decay properties of $|\Xi
\frac{1}{2}^-\rangle_1 $ and $|\Xi \frac{1}{2}^-\rangle_3 $ are
studied as well. Considering the uncertainties of the mass
predictions, we vary the masses of $|\Xi \frac{3}{2}^-\rangle_1$ and
$|\Xi \frac{3}{2}^-\rangle_3$ from 1860 MeV to 1980 MeV. The results
are shown in Fig.~\ref{fig-1690}. From the figure, we find that
$|\Xi \frac{1}{2}^-\rangle_1 $ is a narrow state with a width of
$\Gamma\sim 25$ MeV, while $|\Xi \frac{1}{2}^-\rangle_3 $ is a broad
state with a width of $\sim 100$ MeV. The main decay channels of
$|\Xi \frac{1}{2}^-\rangle_1 $ are $\Sigma\bar{K}$ and
$\Lambda\bar{K}$, while the decays of $|\Xi \frac{1}{2}^-\rangle_3 $
are governed by $\Xi\pi$ and $\Lambda\bar{K}$.

\subsection{$\Xi(1820)$}

In 1987, Biagi \emph{et al.} measured the spin-parity of
$\Xi(1820)$~\cite{Biagi:1986vs}. They found that its spin-parity is
consistent with $J=3/2^-$, which is in good agreement with the quark
model predictions~\cite{Chao:1980em,Melde:2008yr, Pervin:2007wa}. In
the present work, the study of the well-established resonance
$\Xi(1820)$, on the one hand, can provide an important test of our
model; on the other hand, it can let us obtain more information
about the nature of $\Xi(1820)$.

Assigning $\Xi(1820)$ to the negative states with $J^P=1/2^-$,
$3/2^-$ and $5/2^-$, respectively, their strong decays are
calculated with our model. The results are listed in
Tab.~\ref{1820-1}, where we find that only the
$|70,^{2}8,1,1,\frac{3}{2}^-\rangle$ can be assigned to $\Xi(1820)$.
The detail comparisons of theoretical predictions with measurements
are shown in Tab.~\ref{1820b}, from which it is seen that both the
decay width and the partial decay ratios are in agreement with the
measurements of ALITTI 69~\cite{Alitti:1969rb}. Our predictions
about the spin-parity values of $\Xi(1820)$ are consistent with the
other model predictions and experimental determinations.

\begin{table}[ht]
\caption{The decay widths (MeV) of $\Xi(1820)^0$ with different
assignments.} \label{1820-1}
\begin{tabular}{|c|c|c|c|c|c|c|c|c|c|c|c|c }\hline\hline
assignment &~~$~~\Xi\pi$~~&~~$\Sigma \bar{K}  $~~&~~$\Lambda \bar{K} $ &$\Xi(1530)\pi$~~&~~$\Gamma_{total}$\\
\hline
$|70,^{2}8,1,1,\frac{3}{2}^-\rangle$                  &2.85  &10.63        &7.25 &11.98 & 32.71 \\
\hline
$|70,^{4}8,1,1,\frac{1}{2}^-\rangle$                   &64.80  & 23.94        & 20.56  & 0.36& 109.66 \\
$|70,^{4}8,1,1,\frac{3}{2}^-\rangle$                  &4.56 &  0.26           & 0.72 &11.95& 17.49 \\
$|70,^{4}8,1,1,\frac{5}{2}^-\rangle$                  &27.40 & 1.59          &4.35  &1.51 &34.85\\
\hline
$|70,^{2}10,1,1,\frac{1}{2}^-\rangle$                  &4.05  &  5.98       & 5.14 &1.43 &16.60 \\
$|70,^{2}10,1,1,\frac{3}{2}^-\rangle$                  &2.85  & 0.66         &1.81   &11.98 &17.31 \\
\hline
\end{tabular}
\end{table}

\begin{table}[ht]
\caption{The predicted total and partial decay width widths (MeV)
and partial decay width ratios of $\Xi(1820)$ as the
$|70,^{2}8,1,1,\frac{3}{2}^-\rangle$ assignment compared with the
experimental data of ALITTI 69~\cite{Alitti:1969rb}. } \label{1820b}
\begin{tabular}{|c|c|c|c|c|c|c|c|c|c|c }\hline\hline
channel~&~$\Gamma^{th}_i$~&~~$\Gamma^{th}_{total}$
~&~$\Gamma^{exp}_{total}$
& $\frac{\Gamma_i}{\Gamma_{total}}\big|_{th}$&$\frac{\Gamma_i}{\Gamma_{total}}\big|_{exp}$ \\
\hline
$\Xi\pi$ &2.9&32.8&$24^{+15}_{-10}$  &0.09&$0.1\pm0.1$\\
$\Sigma \bar{K}$   &10.6 && &0.32 &$0.30\pm0.15$\\
$\Lambda \bar{K}$ &7.3 && &0.22 &$0.30\pm0.15$\\
$\Xi(1530)\pi$&12.0&& &0.37 &$0.30\pm0.15$\\
\hline
\end{tabular}
\end{table}

Although by considering $\Xi(1820)$ as a pure state
$|70,^{2}8,1,1,\frac{3}{2}^-\rangle$, we note that the theoretical
predictions are in good agreement with the experimental
observations, there still exists room for configuration mixing in
$\Xi(1820)$ for the uncertainties of the experimental data. Chao
\emph{et al.} employed a quark model to study the mass spectrum of
$\Xi$ resonances~\cite{Chao:1980em}. According to their study,
$\Xi(1820)$ should be a mixed state, which dominates by the octet
$|70,^28,1,1,\frac{3}{2}^-\rangle$ components ($\sim 90\%$), while
contains small components of $|70,^210,1,1,\frac{3}{2}^-\rangle$
($\sim 9\%$) and $|70,^48,1,1,\frac{3}{2}^-\rangle$ ($\sim 1\%$).
Thus, we consider the physical state $\Xi(1820)$ as an admixture
between the octet and decuplet states with $J^P$=$\frac{3}{2}^-$.
Using the CKM matrix method discussed in \ref{mix-16}, and fitting
the observed strong decay properties, we obtain
\begin{equation}\label{mix18}
\left(\begin{array}{c}|\Xi \frac{3}{2}^-\rangle_1 \cr |\Xi(1820)
\frac{3}{2}^-\rangle \cr |\Xi \frac{3}{2}^-\rangle_3
\end{array}\right)=\left(\begin{array}{ccc} -0.08 & -0.98& 0.17\cr
 0.96 & -0.13 & -0.26 \cr
 0.27 & 0.14 &0.95
\end{array}\right)\left(\begin{array}{c} |70,^28\rangle \cr |70,^48\rangle \cr |70,^210\rangle
\end{array}\right).
\end{equation}
The theoretical results compared with the data are listed in
Tab.~\ref{1820-2}, where we find that the predicted partial decay
width ratios are in agreement with the experiment data of ALITTI
69~\cite{Alitti:1969rb} as well. Considering configuration mixing
effects in $\Xi(1820)$, the decay width is closer to the center
values of observations. From the mixing parameters obtained in
Eq.~(\ref{mix18}), it is seen that the main component of $\Xi(1820)$
is $|70,^28,1,1,\frac{3}{2}^-\rangle$, which is about $92\%$. As a
mixed state, $\Xi(1820)$ also contains small components of
$|70,^48,1,1,\frac{3}{2}^-\rangle$ ($\sim 0.02\%$) and
$|70,^210,1,1,\frac{3}{2}^-\rangle$ ($\sim 0.07\%$). Our results are
compatible with the predictions in the quark
model~\cite{Chao:1980em}, and the large $N_c$ QCD
approach~\cite{Schat:2001xr,Goity:2002pu}. It should be mentioned
that $\Xi(1820)$ was also suggested to be a dynamically generated
state with $J^P=3/2^-$~\cite{Kolomeitsev:2003kt,Sarkar:2004jh}.

\begin{table}[ht]
\caption{The predicted total and partial decay widths (MeV) and
partial decay width ratios of $\Xi(1820)$ as the mixed state
$|\Xi(1820)\frac{3}{2}^-\rangle$ with mixing angles
$\theta_{12}=85^0$, $\theta_{13}=170^0$ and $\theta_{23}=165^0$. For
a comparison, the experimental data of ALITTI
69~\cite{Alitti:1969rb} are listed as well.} \label{1820-2}
\begin{tabular}{|c|c|c|c|c|c|c|c|c|c|c }\hline\hline
channel~&~$\Gamma^{th}_i$~&~~$\Gamma^{th}_{total}$
~&~$\Gamma^{exp}_{total}$~
 & $\frac{\Gamma_i}{\Gamma_{total}}\big|_{th}$&$\frac{\Gamma_i}{\Gamma_{total}}\big|_{exp}$ \\
\hline
$\Xi\pi$ &2.1&23.3&$24^{+15}_{-10}$  &0.09&$0.1\pm0.1$\\
$\Sigma \bar{K}$   &8.0 && &0.34 &$0.30\pm0.15$\\
$\Lambda \bar{K}$ &9.2 &&&0.40 &$0.30\pm0.15$\\
$\Xi(1530)\pi$&4.0&& &0.17 &$0.30\pm0.15$\\
\hline
\end{tabular}
\end{table}

In brief, $\Xi(1820)$ could be approximately taken as a pure
$|70,^28,1,1,\frac{3}{2}^-\rangle$ state. Slight configuration
mixing may exist in it. All the experimental observations of
$\Xi(1820)$ could be well understood in the constituent quark model.

\begin{figure}[ht]
\centering \epsfxsize=8.5cm \epsfbox{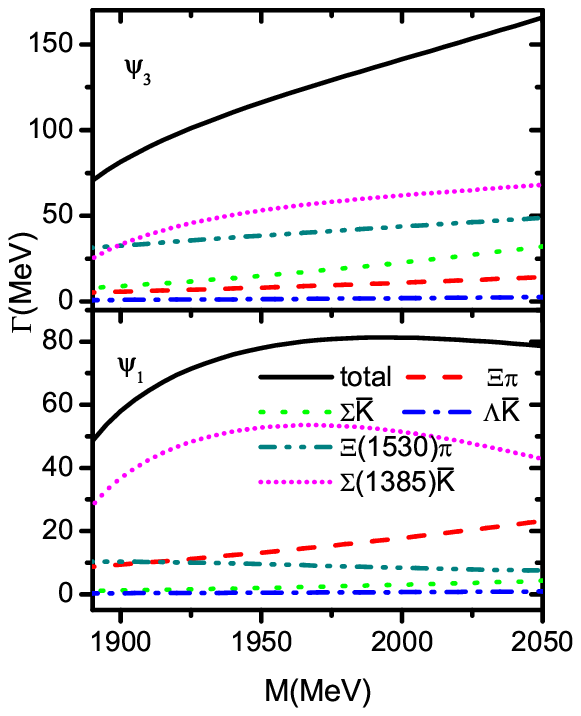} \caption{The decay
properties of $|\Xi \frac{3}{2}^-\rangle_1$ and $|\Xi
\frac{3}{2}^-\rangle_3 $, respectively, where $\Psi_1\equiv|\Xi
\frac{3}{2}^-\rangle_1 $ and $\Psi_3\equiv|\Xi
\frac{3}{2}^-\rangle_3$.}\label{fig-18}
\end{figure}

If $\Xi(1820)$ is a mixed state indeed, its counterparts $|\Xi
\frac{3}{2}^-\rangle_1$ and $|\Xi \frac{3}{2}^-\rangle_3$ in
Eq.~(\ref{mix18}) might be observed in experiments as well. From
Eq.~(\ref{mix18}) it is seen that the main components of $|\Xi
\frac{3}{2}^-\rangle_1$ and $|\Xi \frac{3}{2}^-\rangle_3$ are
$|70,^48,1,1,3/2^-\rangle$ ($\sim 96\%$) and
$|70,^210,1,1,3/2^-\rangle$ ($\sim 90\%$), respectively. According
to the quark model predictions~\cite{Isgur:1978xj,Chao:1980em}, the
masses of $|\Xi \frac{3}{2}^-\rangle_1$ and $|\Xi
\frac{3}{2}^-\rangle_3$ are $\sim 1910$ MeV and $\sim 1970$ MeV,
respectively. The later large QCD
calculations~\cite{Schat:2001xr,Goity:2002pu} gave similar
predictions to those of quark models. The strong decay properties of
$|\Xi \frac{3}{2}^-\rangle_1$ and $|\Xi \frac{3}{2}^-\rangle_3$ are
studied as well. The results are shown in Fig.~\ref{fig-18}. For the
uncertainties of the mass predictions, we vary the masses of $|\Xi
\frac{3}{2}^-\rangle_1$ and $|\Xi \frac{3}{2}^-\rangle_3$ from 1850
MeV to 2050 MeV. From the calculations, we find that if the masses
of these two states are larger than the threshold of $\Sigma(1385)
\bar{K}$, the decay channel $\Sigma(1385) \bar{K}$ dominates their
decays. Furthermore, $\Xi(1530)\pi$ contributes significantly to the
strong decays of these two states.

Finally, it should be pointed out that although the predicted masses
of $|\Xi \frac{3}{2}^-\rangle_1$ and $|\Xi \frac{3}{2}^-\rangle_3$
are close to that of $\Xi(1950)$, the decay modes and partial decay
width ratios are not consistent with the observations. Thus,
$\Xi(1950)$ does not favor any $J^P=3/2^-$ assignments.

\subsection{$\Xi(1950)$}

\begin{figure}[ht]
\centering \epsfxsize=8.8cm \epsfbox{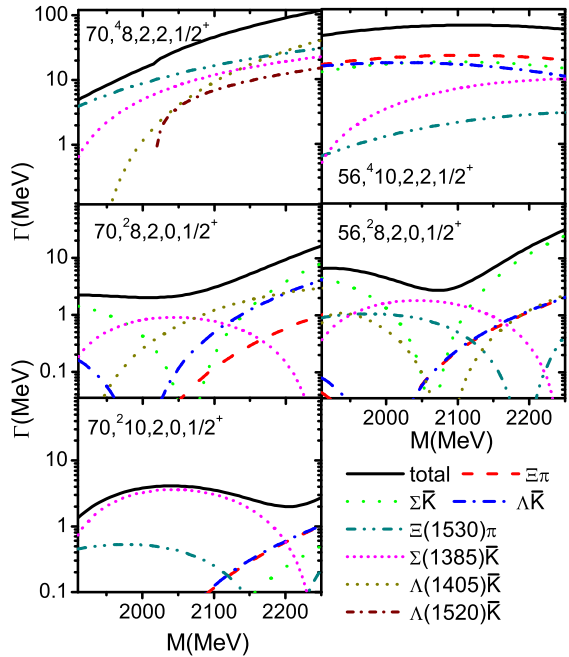} \caption{Strong decay
properties of the $J^P=1/2^+$ excitations in the $N=2$
shell.}\label{fig-1z}
\end{figure}

\begin{figure}[ht]
\centering \epsfxsize=8.8cm \epsfbox{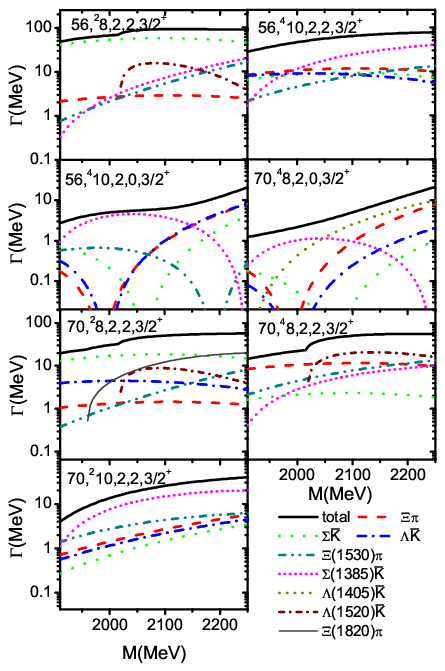} \caption{Strong decay
properties of the $J^P=3/2^+$ excitations in the $N=2$ shell.
}\label{fig-3z}
\end{figure}

\begin{figure}[ht]
\centering \epsfxsize=8.8cm \epsfbox{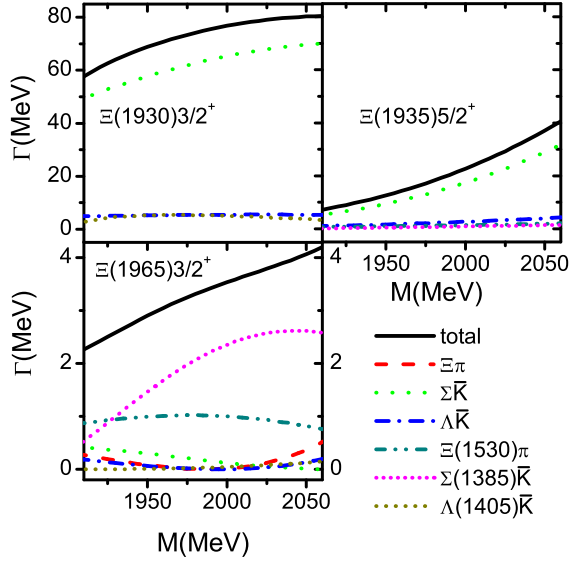} \caption{Strong
decay properties of $J^P=3/2^+$ and $J^P=5/2^+$ mixed states
suggested by Chao \emph{et al.}\cite{Chao:1980em}.}\label{fig-1950}
\end{figure}

\begin{figure}[ht]
\centering \epsfxsize=8.8cm \epsfbox{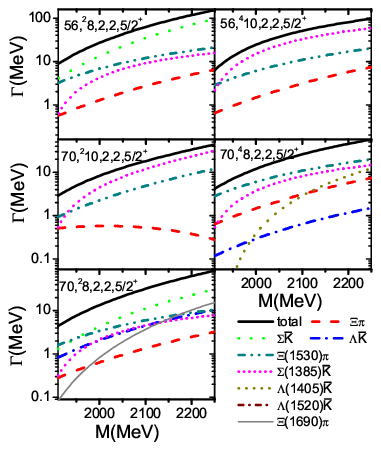} \caption{Strong decay
properties of the $J^P=5/2^+$ excitations in the $N=2$
shell.}\label{fig-5z}
\end{figure}

\begin{figure}[ht]
\centering \epsfxsize=8.0cm \epsfbox{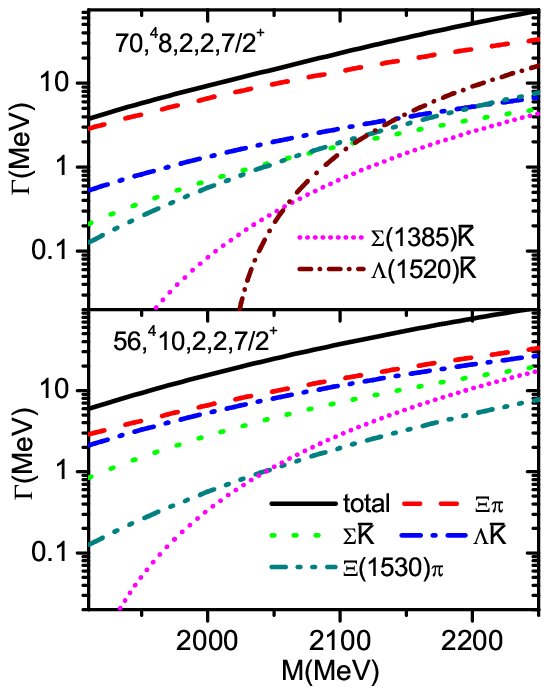} \caption{Strong decay
properties of the $J^P=7/2^+$ excitations in the $N=2$
shell.}\label{fig-7z}
\end{figure}

$\Xi(1950)$ was first observed by Badier \emph{et al.} in the
invariant mass distribution $\Xi\pi$ of $K^-p\rightarrow \Xi^-K\pi$
process~\cite{Badier:1965zzc}, its observed mass and width are
$1933\pm 16$ MeV and $\Gamma\simeq 140\pm 35$ MeV, respectively.
Later, several experimental groups also found structures with a mass
of $\sim 1950\pm 50$ MeV in the other processes. $\Xi(1950)$ was
only observed in $\Xi\pi$, $\Xi(1530)\pi$ and $\Lambda \bar{K}$
decay channels~\cite{Beringer:1900zz}. Most of these structures were
observed in the $\Xi\pi$ channel. In the $\Lambda \bar{K}$ channel,
only Biagi \emph{et al.} claimed that they observed one narrow
structure ($\Gamma\simeq 25\pm 15$ MeV) with a mass of $M\simeq
1963$ MeV~\cite{Biagi:1986vs}. They estimated an upper limit on the
ratio of partial widths $\Gamma(\Sigma \bar{K})/\Gamma(\Lambda
\bar{K})$ of 2.3, and also suggested that the spin-parity of this
resonance should be $5/2^+$ or its spin should be greater than $5/2$
in the natural spin-parity series $7/2^-$, $9/2^+$, etc. While in
the $\Xi(1530)\pi$ channel, only Briefel \emph{et al.} reported that
they observed a broad structure ($\Gamma\simeq 60\pm 39$ MeV) with a
mass of $M\simeq 1964$ MeV~\cite{Briefel:1977bp}. Although
$\Xi(1950)$ is a three-star $\Xi$ resonance listed in PDG, not much
can be said about its properties. According to various model
predictions, there are several $\Xi$ resonances in the $1900-2000$
MeV
region~\cite{Chao:1980em,Schat:2001xr,Goity:2002pu,Goity:2003ab}.

\subsubsection{ $J^P=5/2^-$ assignment }

$\Xi(1950)$ was first classified as the pure octet $\Xi$ resonance
with $J^P=5/2^-$ by Alitti \emph{et al.}~\cite{Alitti:1968zz}. In
1968, they observed a $\Xi$ resonance whose mass and width are
$M=1930\pm 20$ MeV and $\Gamma=80\pm 40$ MeV, respectively. The
resonance parameters were very close to the observations of Badier
\emph{et al.} in 1965~\cite{Badier:1965zzc}. The Gell-Mann-Okubo
mass formula indicates this state might be the pure $J^P=5/2^-$
octet $\Xi$ resonance~\cite{Samios:1974tw,Guzey:2005vz}. The
detailed SU(3) study of the total and partial decay widths of the
$J^P=5/2^-$ octet baryons seemed to give a reasonable and consistent
picture~\cite{Alitti:1968zz}. This classification was also supported
by some studies of the mass spectrum of $\Xi$ resonances in various
quark models~\cite{Chao:1980em,Pervin:2007wa}.

With $\Xi(1950)$ as an assignment to the $J^P=5/2^-$ octet $\Xi$
resonance, we study its strong decay properties, the results are
listed in Tab.~\ref{1950t}. It is seen that the $J^P=5/2^-$
assignment is a broad state with a width of $\sim 100$ MeV. Its
strong decays are dominated by the $\Xi\pi$ channel. The partial
decay widths of $\Xi(1530)\pi$ and $\Lambda \bar{K}$ are sizeable.
Our predictions are in compatible with those
in~\cite{Alitti:1968zz}. Thus, the broad $\Xi$ resonances observed
in $\Xi\pi$ channel might be good candidates for the $J^P=5/2^-$
octet state.

\begin{table}[ht]
\caption{The total and partial decay widths (MeV) of the
well-established four-star baryons $N(1675)$, $\Sigma(1775)$ and
$\Lambda(1830)$, which correspond to
$|70,^{4}8,1,1,\frac{5}{2}^-\rangle$. The data are obtained from
PDG.} \label{1950t}
\begin{tabular}{|c|c|c|c|c|c|c|c|c|c|c }\hline\hline
assignment &channel~&$\Gamma^{th}_i $~~&~$\Gamma^{th}_{total}$
&$\Gamma^{exp}_{total}$
& $\frac{\Gamma_i}{\Gamma_{total}}\big|_{th}$&$\frac{\Gamma_i}{\Gamma_{total}}\big|_{exp}$\\
\hline
$N(1675)\frac{5}{2}^-$                  &$n\pi$ &  25.3         &81 & $130\sim165$& 0.31 & 0.35--0.45\\
                   &$\Delta\pi$ &50.6   &    &&0.62& 0.50--0.60\\
                   &$N \eta$ &  5.8  &    & &0.07& 0.01\\
\hline
$\Sigma(1775)\frac{5}{2}^-$                  &$\Sigma\pi$ &  8.0          &76&$105\sim135$& 0.11&0.02--0.05 \\
                  &$\Lambda\pi^0$  & 19.1         & & &0.25& 0.14--0.20\\
                   &$n\bar{K}$  &32.1          & &  &0.42&0.37--0.43\\
                   &$\Sigma(1385)\pi$&4.8&& &0.06&0.08--0.12\\
                   &$\Delta(1232)\bar{K}$&1.0&& &0.01&...\\
                   &$\Lambda(1520)\pi$&11.2&& &0.14&0.17--0.23\\
\hline
$\Lambda(1830)\frac{5}{2}^-$                 &$\Sigma\pi$ &  52.2         &94 & $60\sim110$&0.55&0.35--0.75 \\
                  &$\Sigma(1385)\pi$ &  41.7        &    & &0.45&$>0.15$\\
\hline
$\Xi(1950)\frac{5}{2}^-$       &$\Xi\pi$              &  71.33         &105 & $80\pm40$&0.68& ...\\
                  &$\Sigma\bar{K}$       &  7.8          &    & &0.07& ... \\
                  &$\Lambda\bar{K}$      &  13.6        &    & &0.13& ... \\
                  &$\Xi(1530)\pi$        &  10.9        &    & &0.10& ... \\
                  &$\Sigma(1385)\bar{K}$ &  1.2         &    & &0.01& ... \\
\hline
\end{tabular}
\end{table}

As a by-product, we calculate the strong decays of the other members
of $J^P=5/2^-$ octet baryons, $N(1675)\frac{5}{2}^-$,
$\Sigma(1775)\frac{5}{2}^-$ and $\Lambda(1830)\frac{5}{2}^-$. The
results are listed in Tab.~\ref{1950t} as well. From the table, it
is found that our predictions of the strong decay properties of
$N(1675)\frac{5}{2}^-$, $\Sigma(1775)\frac{5}{2}^-$ and
$\Lambda(1830)\frac{5}{2}^-$ are in reasonable agreement with the
observations.

\subsubsection{ $J^P=1/2^-$ assignment}

Recently, Valderrama, Xie and Nieves proposed the existence of a
spin-parity state $J^P=1/2^-$ decuplet belonging to
$\Xi(1950)$~\cite{PavonValderrama:2011gp}. Now we discuss the
possibilities of $\Xi(1950)$ as an assignment to the $J^P=1/2^-$
states. In Sec.~\ref{mix-16}, we predicted $\Xi(1690)$ is a mixed
state with $J^P=1/2^-$. According to the mass calculations of
constituent quark models~\cite{Chao:1980em,Chen:2009de} and Large
$N_c$ QCD~\cite{Schat:2001xr,Goity:2002pu}, the masses of the
counterparts of $\Xi(1690)$, $|\Xi \frac{1}{2}^-\rangle_1$ and $|\Xi
\frac{1}{2}^-\rangle_3 $, might be close to 1950 MeV. Thus, they
might be candidates for $\Xi(1950)$. The strong decay properties of
$|\Xi \frac{1}{2}^-\rangle_1$ and $|\Xi \frac{1}{2}^-\rangle_3 $ had
been studied in Sec.~\ref{mix-16} (see Fig.~\ref{fig-1690}).
Considering $|\Xi \frac{1}{2}^-\rangle_1$ as an assignment to
$\Xi(1950)$, both the decay width and partial width ratio,
\begin{equation}
\Gamma\simeq 27~ \mathrm{MeV},
\frac{\Gamma(\Sigma\bar{K})}{\Gamma(\Lambda \bar{K})}\simeq 2.0,
\end{equation}
are in agreement with the observations of Biagi
87C~\cite{Biagi:1986vs}; however, the spin-parity $J^P=1/2^-$
disagrees with their suggestion. While assigning $|\Xi
\frac{1}{2}^-\rangle_3 $ to $\Xi(1950)$, we note that its width and
partial width ratios
\begin{equation}
\Gamma\simeq 84~ \mathrm{MeV},
\frac{\Gamma(\Sigma\bar{K})}{\Gamma(\Lambda \bar{K})}\simeq
1.6,~\frac{\Gamma(\Xi\pi)}{\Gamma(\Sigma\bar{K})}\simeq 2.3,
\end{equation}
are consistent with those of the broad structures observed in the
$\Xi\pi$ channel. Thus, the spin-parity $1/2^-$ mixed state $|\Xi
\frac{1}{2}^-\rangle_3 $ could be a good assignment to $\Xi(1950)$.

\subsubsection{ $J^P=1/2^+$ assignment }

In Ref.~\cite{Oh:2007cr}, Oh predicted that $\Xi(1950)$ might have
$J^P=1/2^+$ in the Skyrme model. We calculate the strong decays of
all the excitations of $\Xi$ with $J^P=1/2^+$ in the $N=2$ shell.
The results are shown in Fig.~\ref{fig-1z}. From the figure, it is
found that if the second orbital excitation
$|56,^{4}10,2,2,\frac{1}{2}^+\rangle$  is considered as an
assignment to $\Xi(1950)$, the decay width and partial width ratios
are
\begin{equation}
\Gamma\simeq 53~ \mathrm{MeV},
\frac{\Gamma(\Sigma\bar{K})}{\Gamma(\Lambda \bar{K})}\simeq
0.88,~\frac{\Gamma(\Xi\pi)}{\Gamma(\Sigma\bar{K})}\simeq 1.27.
\end{equation}
The decay width and decay modes are consistent with the observations
of Goldwasser \emph{et al.}~\cite{Goldwasser:1970fk}. It should be
pointed out that the predicted mass of various models for
$|56,^{4}10,2,2,\frac{1}{2}^+\rangle$ is $\sim2$ GeV
~\cite{Chao:1980em,Goity:2003ab,Chen:2009de}, which is slightly
larger than the observation.

\subsubsection{ $J^P=3/2^+$ assignment }

Furthermore, we calculate the strong decays of all the excitations
of $\Xi$ with $J^P=3/2^+$ in the $N=2$ shell. The results are shown
in Fig.~\ref{fig-3z}. It is found that the second orbital excitation
$|56,^{4}10,2,2,\frac{3}{2}^+\rangle$ might be a candidate for
$\Xi(1950)$. The decay width
\begin{equation}
\Gamma\simeq 36~ \mathrm{MeV},
\end{equation}
and the partial width ratios
\begin{equation}
\frac{\Gamma(\Sigma\bar{K})}{\Gamma(\Lambda \bar{K})}\simeq
0.88,~\frac{\Gamma(\Xi\pi)}{\Gamma(\Sigma\bar{K})}\simeq
1.27,~~\frac{\Gamma(\Xi\pi)}{\Gamma(\Xi(1530)\pi)}\simeq 3.16
\end{equation}
are in agreement with the observations of Biagi
87C~\cite{Biagi:1986vs} and APSELL 70~\cite{Apsell:1970uf}. However,
the spin-parity of $|56,^{4}10,2,2,\frac{3}{2}^+\rangle$ is not
consistent with the moment analysis of Biagi
87C~\cite{Biagi:1986vs}. Furthermore, the predicted mass of
$|56,^{4}10,2,2,\frac{3}{2}^+\rangle$ in large $N_c$ QCD is
obviously larger than that of $\Xi(1950)$~\cite{Goity:2003ab}.

Including configuration mixing effects, Chao \emph{et al.} predicted
two $J^P=3/2^+$ mixed states with masses of $M=1930$ and 1965 MeV,
respectively~\cite{Chao:1980em}. With their mixing scheme, we
predict the strong decays of the two states. The results are shown
in Fig.~\ref{fig-1950}. It is obviously seen that the decay
properties of these mixed states are not in agreement with any
observations of $\Xi(1950)$.

\begin{table}[ht]
\caption{The total and partial decay widths (MeV) of the
well-established four-star baryons $N(1680)$, $\Sigma(1915)$ and
$\Lambda(1820)$, which are considered the pure
$|56,^{2}8,2,2,\frac{5}{2}^+\rangle$ state. The data are obtained
from PDG.} \label{2fenz5}
\begin{tabular}{|c|c|c|c|c|c|c|c|c|c|c }\hline\hline
assignment &channel~&$\Gamma^{th}_i $~~&~$\Gamma^{th}_{total}$
&$\Gamma^{exp}_{total}$
& $\frac{\Gamma_i}{\Gamma_{total}}\big|_{th}$&$\frac{\Gamma_i}{\Gamma_{total}}\big|_{exp}$\\
\hline
$N(1680)\frac{5}{2}^+$ &$n\pi$ &  38.0         &59 & $120\sim140$& 0.64 & 0.65--0.70\\
                   &$\Delta\pi$ &21.0   &    &&0.36& ... \\
                   &$N \eta$ &  0.4  &    & &0.01& 0.0--0.01\\
                   &$\Lambda K$ &0.03   &    &&0& ...\\
\hline $\Sigma(1915)\frac{5}{2}^+$
                   &$N\bar{K}$         & 1.2        &63 &$80\sim160$ &0.02& 0.05--0.15\\
                   &$\Lambda\pi$       &10.4          & &  &0.16&... \\
                   &$\Sigma\pi$        &26.7 && &0.42& ...\\
                   &$\Sigma(1385)\pi$  &6.2&& &0.10&... \\
                   &$\Delta(1232)\bar{K}$    &18.4&& &0.29 &...\\
\hline $\Lambda(1820)\frac{5}{2}^+$
                   &$N\bar{K}$         & 14.8         & 30 & $70\sim90$ &0.49& 0.55--0.65\\
                   &$\Sigma\pi$        &5.8&& &0.19&0.08--0.14\\
                   &$\Sigma(1385)\pi$  &9.8&& &0.32&  ...  \\
\hline
$\Xi(1963)\frac{5}{2}^+$       &$\Xi\pi$  &  1.0          &19  & $25\pm15$& 0.05& \\
                  &$\Sigma\bar{K}$        &  9.2          &    &          & 0.48&  \\
                  &$\Lambda\bar{K}$       &  0.8          &    &          & 0.04&  \\
                  &$\Xi(1530)\pi$         & 5.2           &    &          & 0.27&  \\
                  &$\Sigma(1385)\bar{K}$  & 2.5           &    &          & 0.13&  \\
\hline
\end{tabular}
\end{table}

\subsubsection{ $J^P=5/2^+$ assignment}

Recently, Valderrama, Xie and Nieves~\cite{PavonValderrama:2011gp}
predicted that the narrow structure with a mass of $M\simeq 1963$
MeV observed in the $\Lambda \bar{K}$ channel by Biagi \emph{et
al.}~\cite{Biagi:1986vs} (denoted by $\Xi(1963)$) could be assigned
to the partner of the $J^P=5/2^+$ $N(1680)$, $\Lambda(1820)$ and
$\Sigma(1915)$ resonances. According to the classification of quark
model, these resonances could be roughly considered as the
$|56,^{2}8,2,2,\frac{5}{2}^+\rangle$
configuration~\cite{Klempt:2009pi}. Firstly, we calculate the strong
decays of the $N(1680)$, $\Lambda(1820)$ and $\Sigma(1915)$ states,
which are listed in Tab.~\ref{2fenz5}. From the table, it is found
that the decay properties of $N(1680)$, $\Lambda(1820)$ and
$\Sigma(1915)$ could be roughly understood by taking them as the
$|56,^{2}8,2,2,\frac{5}{2}^+\rangle$ configuration. According to the
model predictions, the mass of
$\Xi|56,^{2}8,2,2,\frac{5}{2}^+\rangle$ is close to 1963 MeV
~\cite{Chao:1980em,Goity:2003ab}. Assigning $\Xi(1963)$ to the
$|56,^{2}8,2,2,\frac{5}{2}^+\rangle$ configuration, the predicted
strong decay properties are shown in Tab.~\ref{2fenz5} as well. It
is seen that although the decay width is in agreement with the data,
the partial decay width ratios are not consistent with the
observations of Biagi \emph{et al.}~\cite{Biagi:1986vs} at all.

Furthermore,  Chao \emph{et al.} also predicted a mixed states with
$J^P=5/2^+$ having a mass of $M=1935$ MeV~\cite{Chao:1980em}. With
their mixing scheme, its strong decays are studied. The results are
shown in Fig.~\ref{fig-1950} as well. It is seen that the strong
decays are dominated by the $\Sigma \bar{K}$ channel, which
disagrees with any observations of $\Xi(1950)$.

In brief, there might be several $\Xi$ resonances in the $1900-2000$
MeV region observed in experiments. These states are most likely to
correspond to the pure octet $\Xi$ resonance with $J^P=5/2^-$, the
mixed state $|\Xi \frac{1}{2}^-\rangle_3 $ with $J^P=1/2^-$ (its
main component is $|70,^48,1,1,\frac{1}{2}^-\rangle$), the second
orbital excitation $|56,^{4}10,2,2,\frac{1}{2}^+\rangle$ with
$J^P=1/2^+$, etc.  These states can be easily distinguished by the
measurements of their partial decay width ratios,
$\Gamma(\Lambda\bar{K})/\Gamma(\Sigma\bar{K})$ and
$\Gamma(\Sigma\bar{K})/\Gamma(\Xi\pi)$.

\begin{widetext}
\begin{center}
\begin{table}[ht]
\caption{The decay widths (MeV) of $\Xi(2030)$ with different
assignments.} \label{T2030}
\begin{tabular}{|c|c|c|c|c|c|c|c|c|c|c|c|c }\hline\hline
assignment &~$~\Xi\pi$~&~$\Sigma \bar{K} $~&~$\Lambda \bar{K}  $~&~
$\Xi(1530)\pi $~&~$\Sigma(1385)\bar{K}$~
&~$\Lambda(1405)\bar{K}$~&~$\Lambda(1520)\bar{K}$ &~$\Gamma_{total}$& $\Gamma(\Lambda\bar{K}):\Gamma(\Sigma\bar{K})$\\
\hline
$|56,^{2}8,2,2,\frac{5}{2}^+\rangle$  &1.6  &18.5 &1.3  &8.2  &5.7  &0.4 &0.01&35.8&0.07\\
\hline
$|70,^{2}8,2,2,\frac{3}{2}^+\rangle$                 &1.4 &17.9   &4.5   & 1.4   &0.6 &1.5&5.4&32.7&0.25\\
$|70,^{2}8,2,2,\frac{5}{2}^+\rangle$                 &0.8  & 5.8   &2.7   &4.1    &2.8 &0.05  &0.07 &16.3&0.46\\
\hline
\end{tabular}
\end{table}
\end{center}
\end{widetext}

\subsection{$\Xi(2030)$}

$\Xi(2030)$ is a three-star state listed in PDG. It mainly decays
into $\Sigma\bar{K}$ and $\Lambda\bar{K}$ channels. The decay ratios
into the other channels, such as $\Xi\pi$ and $\Xi(1530)\pi$, are
small. The measured decay width and partial decay width ratio are
\begin{equation}
\Gamma\simeq 21\pm 6
\mathrm{MeV},~\frac{\Gamma(\Lambda\bar{K})}{\Gamma(\Sigma
\bar{K})}\simeq0.22\pm 0.09.
\end{equation}
A moment analysis of the HEMINGWAY 77 data indicated at a level of
three standard deviations that $J\geq 5/2$~\cite{Hemingway:1977uw}.

We analyze the strong decay properties for all the configurations in
the $N=2$ shell, which are shown in Figs.~\ref{fig-1z}-\ref{fig-7z}.
From the figures, we find that only three excitations
$|56,^{2}8,2,2,\frac{5}{2}^+\rangle$,
$|70,^{2}8,2,2,\frac{3}{2}^+\rangle$ and
$|70,^{2}8,2,2,\frac{5}{2}^+\rangle$ have comparable decay widths
with that of $\Xi(2030)$, and mainly decay into $\Sigma\bar{K}$
channel. Considering them as assignments to $\Xi(2030)$, we collect
their decay properties in Tab.~\ref{T2030}, where we find that both
the decay width and the partial decay ratio
\begin{equation}
\Gamma\simeq
33~\mathrm{MeV},~\frac{\Gamma(\Lambda\bar{K})}{\Gamma(\Sigma
\bar{K})}\simeq0.25.
\end{equation}
of the $|70,^{2}8,2,2,\frac{3}{2}^+\rangle$ configuration are in
good agreement with the observations. However, its spin $J=3/2$
disagrees with the moment analysis of the HEMINGWAY 77
data~\cite{Hemingway:1977uw}.

As a whole, if we do not care about the moment analysis of the
HEMINGWAY 77 data~\cite{Hemingway:1977uw}, $\Xi(2030)$ favors the
$|70,^{2}8,2,2,\frac{3}{2}^+\rangle$ assignment. $\Xi(2030)$ could
not be assigned to any pure $J^P=7/2^+$ states or any admixtures
between them. If the spin-parity of $\Xi(2030)$ is $J^P=5/2^+$, it
is most likely to be a mixed state, for no pure $J^P=5/2^+$
configuration could explain the data.

\section{Summary}\label{sum}

In this work, we have studied the strong decays of the $\Xi$ baryons
within $N\leq 2$ shells in a chiral quark model. The strong decay
properties of these well-established ground decuplet baryons could
be reasonably described. We find that $\Xi(1690)$ should be assigned
to the spin-parity $J^P=1/2^-$ state
$|70,^{2}8,1,1,\frac{1}{2}^-\rangle$, which might slightly mix with
the other configurations. The $J^P=1/2^-$ for $\Xi(1690)$ predicted
by us is consistent with the suggestions from the constituent quark
model~\cite{Pervin:2007wa}, Skyrme model~\cite{Oh:2007cr} and
unitary chiral approaches~\cite{Kolomeitsev:2003kt,Sarkar:2004jh}.
The strong decays of the physical partners of $\Xi(1690)$, $|\Xi
\frac{1}{2}^-\rangle_1 $ and $|\Xi \frac{1}{2}^-\rangle_3 $ are
analyzed as well. $|\Xi \frac{1}{2}^-\rangle_1 $ might be observed
in the $\Sigma\bar{K}$ and $\Lambda\bar{K}$ channels, while $|\Xi
\frac{1}{2}^-\rangle_3 $ is possibly observed in the $\Xi\pi$ and
$\Lambda\bar{K}$ channels.

The strong decay properties of $\Xi(1820)$ could be well understood
by assigning it to $|70,^{2}8,1,1,\frac{3}{2}^-\rangle$. There might
exist slight configuration mixing in $\Xi(1820)$. Its main component
is $|70,^{2}8,1,1,\frac{3}{2}^-\rangle$ ($\sim 92\%$), which is
compatible with the predictions of the quark
model~\cite{Chao:1980em}, and large $N_c$ QCD approach
~\cite{Schat:2001xr,Goity:2002pu}. Considering the configuration
mixing effects, we also have studied the strong decay properties of
the physical partners of $\Xi(1820)$, $|\Xi \frac{3}{2}^-\rangle_1$
and $|\Xi \frac{3}{2}^-\rangle_3 $. The observations in the
$\Sigma(1385)\bar{K}$ and $\Xi(1530)\pi$ channels are crucial to
look for the other $J^P=3/2^-$ states $|\Xi \frac{3}{2}^-\rangle_1$
and $|\Xi \frac{3}{2}^-\rangle_3 $ in future experiments.

The situation for $\Xi(1950)$ is very complicated. Several $\Xi$
resonances in the $1900-2000$ MeV region might have been observed in
experiments, which is supported by the mass calculations from
various
models~\cite{Chao:1980em,Schat:2001xr,Goity:2002pu,Goity:2003ab},
and the recent strong decay analysis in
~\cite{PavonValderrama:2011gp}. The broad $\Xi$ resonances observed
in the $\Xi\pi$ channel might be good candidates for the $J^P=5/2^-$
octet state or the mixed state $|\Xi \frac{1}{2}^-\rangle_3 $ with
$J^P=1/2^-$. The $\Xi$ resonance with moderate width observed by
Goldwasser \emph{et al.}~\cite{Goldwasser:1970fk} might correspond
to the $J^P=1/2^+$ excitation $|56,^{4}10,2,2,\frac{1}{2}^+\rangle$.
The existence of a $J^P=1/2^+$ $\Xi$ excitation with a mass around
1950 MeV is also suggested by Oh in a Skyrme model~\cite{Oh:2007cr}.
The second orbital excitation $|56,^{4}10,2,2,\frac{3}{2}^+\rangle$
and the mixed state $|\Xi \frac{1}{2}^-\rangle_1$ might be
candidates for a narrow width state observed in the $\Lambda
\bar{K}$ channel; however, their spin-parity is not consistent with
a moment analysis of the data. The partial decay ratios,
$\Gamma(\Lambda\bar{K})/\Gamma(\Sigma\bar{K})$ and
$\Gamma(\Sigma\bar{K})/\Gamma(\Xi\pi)$, are sensitive to different
assignments. Thus, the measurements of these ratios are crucial to
uncover many puzzles in $\Xi(1950)$.

In present work, $\Xi(2030)$ as any spin-parity $J^P=7/2^+$ states
should be excluded. The observations of $\Xi(2030)$ do not favor any
pure $J^P=5/2^+$ configuration as well. If we do not care about the
moment analysis of the HEMINGWAY 77 data, $\Xi(2030)$ favors the
$|70,^{2}8,2,2,\frac{3}{2}^+\rangle$ assignment. Further
observations in the $\Xi(1530)\pi$ and $\Sigma(1385) \bar{K}$
channels are necessary.

To provide helpful information for the search for the missing $\Xi$
baryons, in Figs.~\ref{fig-1z}-\ref{fig-7z} our predictions of their
strong decay properties are shown as well. From our theoretical
results, we find that the strong decays of many $\Xi$ resonances are
dominated by the $\Xi(1530)\pi$ and $\Sigma(1385) \bar{K}$, thus, in
these decay channels, we might find some new $\Xi$ resonances as
well.

\section*{ Acknowledgements }

This work is supported, in part, by the National Natural Science
Foundation of China (Grant No. 11075051), Program for Changjiang
Scholars and Innovative Research Team in University (Grant No.
IRT0964), the Program Excellent Talent Hunan Normal University, and
the Hunan Provincial Natural Science Foundation (Grants No.11JJ7001
and No. 13JJ1018).


\end{document}